\documentstyle[multicol,eqsecnum,psfig,epsf,aps,feynmf]{revtex}
\newcommand{\la}{\left\langle}
\newcommand{\ra}{\right\rangle}
\newcommand{\be}{\begin{equation}}
\newcommand{\ee}{\end{equation}}
\newcommand{\bea}{\begin{eqnarray}}
\newcommand{\eea}{\end{eqnarray}}
\newcommand{\ba}{\begin{array}}
\newcommand{\ea}{\end{array}}
\newcommand{\eqnKrai}
{\begin{equation} \label{eqn:Krai65}
E^u(k) = E^b(k) = A (\Pi B_0)^{1/2} k^{-3/2} 
\end{equation}}
\newcommand{\eqnKolmMHD}
{\be \label{eqn:KolmMHD}
E^{\pm}(k) =  K^{\pm} \frac{\left( \Pi^{\pm} \right)^{4/3}}
 		          {\left( \Pi^{\mp} \right)^{2/3}} k^{-5/3}
\ee}
\newcommand{\eqnMHDubk}
{\bea 
\left( -i\omega + \nu k^2 \right) u_i (\hat{k})  & 
 =  & -\frac{i}{2} P^+_{ijm}({\bf k}) \int d\hat{p} 
          [ u_j (\hat{p}) u_m (\hat{k}-\hat{p}) -
           b_j (\hat{p}) b_m (\hat{k}-\hat{p}) ] \label{eqn:udot} \\
\left( -i\omega + \eta k^2 \right) b_i (\hat{k})  & 
 = &  - i P^-_{ijm}({\bf k}) \int d\hat{p} 
          [ u_j (\hat{p}) b_m (\hat{k}-\hat{p}) ]
       		\label{eqn:bdot}   \\
k_i u_i({\bf k}) & = & 0 \\
k_i b_i({\bf k}) & = & 0  \eea}

\begin{document}
\draft
\title{Calculation of renormalized viscosity
and resistivity in magnetohydrodynamic turbulence}
\author{Mahendra\ K.\ Verma \thanks{email: mkv@iitk.ac.in}}
\address{Department of Physics, Indian Institute of Technology,
Kanpur  --  208016, INDIA}
\date{May 2001}
\maketitle

\begin{abstract}
A self-consistent renormalization (RG) scheme has been applied to
nonhelical magnetohydrodynamic turbulence with normalized cross
helicity $\sigma_c =0$ and $\sigma_c \rightarrow 1$.  Kolmogorov's
$5/3$ powerlaw is assumed in order to compute the renormalized
parameters.  It has been shown that the RG fixed point is stable for
$d \ge d_c \approx 2.2$.  The renormalized viscosity $\nu^*$ and
resistivity $\eta^*$ have been calculated, and they are found to be
positive for all parameter regimes.  For $\sigma_c=0$ and large
Alfv\'{e}n ratio (ratio of kinetic and magnetic energies) $r_A$,
$\nu^*=0.36$ and $\eta^*=0.85$.  As $r_A$ is decreased, $\nu^*$
increases and $\eta^*$ decreases, untill $r_A \approx 0.25$ where both
$\nu^*$ and $\eta^*$ are approximately zero.  For large $d$, both
$\nu^*$ and $\eta^*$ vary as $d^{-1/2}$.  The renormalized parameters
for the case $\sigma_c \rightarrow 1$ are also  reported.
\end{abstract}
\vspace{1.5cm}

\section{Introduction}
The Renormalization Group (RG) technique provided an important tool
for understanding  of the phenomena of phase transitions and critical
phenomena.  Motivated by this success,  researchers have applied this
technique to turbulence and other nonequilibrium systems.
In this paper we apply RG to magnetohydrodynamic (MHD) turbulence.

Forster {\em et al.}~\cite{FNS} first applied dynamical RG procedure to the
analysis of Navier-Stokes and Burgers equations, both stirred by
random forces.  They eliminated the large-wavenumber modes and included
the effects of nonlinearity in the effective viscosity.  Later,
DeDominicis and Martin \cite{DeDo}, Fournier and Frisch
\cite{FourFris}, Yakhot and Orszag \cite{YakhOrsz}, many others
studied various aspects of RG for fluid turbulence, but all of them
considered certain form of external forcing.  McComb and his coworkers
(\cite{McCo:book,McCo:rev} and reference therein) instead applied
self-consistent RG procedure; here the energy spectrum was assumed to
be Kolmogorov's spectrum, and the renormalized viscosity was computed
iteratively.  We have adopted a similar scheme for MHD turbulence.

Fournier {\em et al.}~\cite{FourSule}, Camargo and Tasso~\cite{Cama}, and
Liang and Diamond~\cite{Lian} employed RG technique to MHD turbulence
on the similar lines as adopted by Forster {\em et al.}~\cite{FNS} for fluid
turbulence.  Fournier {\em et al.}~\cite{FourSule} found that for $d > d_c
\approx 2.8$, there are two nontrivial regimes: a kinetic regime where
the renormalization of the transport coefficients is due to kinetic
small-scales; and a magnetic regime where it is due to the magnetic
small-scales. In dimensions $2 \le d \le d_c$, there is no stable
fixed point for turbulence with sufficiently strong external currents.
Camargo and Tasso~\cite{Cama} studied the effective (renormalized)
viscosity and resistivity and concluded that negative viscosity and
resistivity are not permissible except in certain cases.  Liang and
Diamond \cite{Lian} showed that no RG fixed point exists in 2D MHD
turbulence.  All the above authors however do not give any clear idea
about the energy spectrum of MHD turbulence, which still remains
controversial.  However, Verma \cite{MKV:B0RG} used technique similar
to that of McComb and Coworkers (\cite{McCo:book,McCo:rev} and references
therein) and showed that Kolmogrov's spectrum is a consistent
solution of MHD RG equations.

Kraichnan \cite{Krai65} and Irosnikov \cite{Iros} first gave
phenomenology of steady-state, homogeneous, and isotropic MHD
turbulence.  They argued that the kinetic and magnetic energy spectra
($E^u(k)$ and $E^b(k)$ respectively) are
\eqnKrai
where $\Pi$ is the total energy cascade rate, $B_0$ is the mean
magnetic field or the magnetic field of the largest eddy, and $A$ is
a universal constant of order one.  Later 
Matthaeus and Zhou \cite{MattZhou}, and Zhou and Matthaeus
\cite{ZhouMatt} proposed a generalized phenomenology in which the
cascade rates $\Pi^{\pm}$ and 
energy spectra $E^{\pm}(k)$ of Els\"{a}sser variables ${\bf z^{\pm} (
= u \pm b})$ are related by
\be
\Pi^{\pm} = \frac{A^2 E^+(k) E^-(k) k^3}{B_0 + \sqrt{k E^{\pm} (k)}}
\ee
where $A$ is a constant.
In the limit $B_0=0$, we can immediately obtain
\eqnKolmMHD
where $K^{\pm}$ are  Kolmogorov's constants for MHD.  The
Eq.~(\ref{eqn:KolmMHD}) was first derived by Marsch \cite{Mars:Kolm}.

The energy spectrum of the solar wind, a well known observational
platform for MHD turbulence, is closer to $k^{-5/3}$ than to
$k^{-3/2}$ \cite{MattGold,MarsTu90}. In addition, the recent numerical
\cite{MKV:mhdsim,Bisk1:Kolm,Bisk2:Kolm} and theoretical
\cite{MKV:B0RG,Srid1,Srid2} work support Kolmogorov-like phenomenology
for MHD turbulence. Curiously, Kolmogorov's spectrum is observed even
in the presence of a mean magnetic field.  To understand
theoretically, Verma \cite{MKV:B0RG}, Sridhar and Goldreich
\cite{Srid1}, and Goldreich and Sridhar \cite{Srid2} applied field
theoretic techniques to MHD turbulence. Verma \cite{MKV:B0RG} showed
that the mean magnetic field $B_0$ responsible for the Alfv\'{e}n
effect gets renormalized in strong turbulence ($B_0(k) \propto
k^{-1/3}$).  When we substitute the renormalized $B_0$ in
Eq.~(\ref{eqn:Krai65}), the energy spectrum turns out to be
proportional to $k^{-5/3}$. Sridhar and Goldreich \cite{Srid1}, and
Goldreich and Sridhar \cite{Srid2} used wave interaction picture and
showed that the energy spectrum is anisotropic, and it follows
Kolmogorov's powerlaw for strong turbulence.

In this paper we calculate the renormalized viscosity and resistivity
for MHD.  Here we test whether Kolmogorov's spectrum is a consistent
solution of RG or not if the turbulence is forced at large-scale.  We
carry out wavenumber elimination and obtain recursive RG
equations. Kolmogorov's spectrum is substituted for the correlation
function in the recursive RG equation.  After that we attempt to
obtain a convergent solution for the renormalized viscosity and
resistivity.  If the convergent solution exists, then the Kolmogorov's
spectrum is one of the correct choices for the energy spectrum of MHD
turbulence.  For these cases we compute many important quantities,
e.g., renormalized viscosity, renormalized resistivity, cascade rates,
etc., in a reasonably simple manner. In this paper we report the
calculation of renormalized viscosity and resistivity, and in the
subsequent paper (referred to as paper II) we will report the cascade
rate calculation.

There are two field variables in MHD: the velocity fluctuation ${\bf
u}$ and the magnetic field fluctuation ${\bf b}$.  The quantities of
interest are Kinetic energy (KE), magnetic energy (ME), total energy
(KE+ME), and cross helicity ($H_c = {\bf u \cdot b}$).  The
dimensionless parameters used in this paper are the ratio of twice the
cross helicity and the total energy, called the normalized cross
helicity $\sigma_c$ ($\sigma_c = 2 H_c / (KE+ME)$), and the ratio of
KE and ME, called the Alfv\'{e}n ratio $r_A$.  In this paper we have
assumed that the mean magnetic field is zero, and so are magnetic and
kinetic helicities, i.e., we are considering nonhelical plasma.  The
calculation of the renormalized parameters is quite complex for
arbitrary $\sigma_c$ and $r_A$.  Therefore, we limit ourselves to two
limiting cases: (1) $\sigma_c=0$ and whole range of $r_A$; (2)
$\sigma_c \rightarrow 1$ and $r_A=1$.  Even though full range of
$\sigma_c$ is observed in terrestrial and extraterrestrial plasmas,
the case $\sigma_c \approx 0$ is most pronounced.

The outline of the paper is as follows: in sections 2 and 3 we compute
the renormalized viscosity and resistivity for the cases $\sigma_c=0$
and $\sigma_c \rightarrow 1$ respectively.  Section 4 contains the
summary and conclusions.

\section{RG Calculation: $\sigma_c=0$}
In this section we will calculate the renormalized viscosity and
resistivity of MHD equations for zero mean magnetic field and zero
cross helicity ($\sigma_c =0$). In this case, the RG equations in
terms of velocity and magnetic field variables are simpler as compared
to those with Els\"{a}sser variables.  Therefore, we will work with
${\bf u}$ and ${\bf b}$ variables.  We take the following form of
Kolmogorov's spectrum for KE [$E^u(k)$] and ME [$E^b(k)$]
\bea 
	E^u(k) & = & K^u \Pi^{2/3} k^{-5/3}, \label{eqn:Euk} \\ 
	E^b(k) & = & E^u(k) / r_A, \label{eqn:Ebk} 
\eea
where $K^u$ is Kolmogorov's constant for MHD turbulence, and $\Pi$ is
the total energy flux.  In the limit $\sigma_c=0$, we have
$E^{+}=E^{-}$ and $\Pi^+ = \Pi^{-} = \Pi$
[cf. Eq.~(\ref{eqn:KolmMHD})].  Therefore, $E_{total}(k) = E^{+}(K) =
E^u(k)+E^b(k)$ and
\be 
K^+ = K^u (1+r_A^{-1}) 
\ee

With this preliminaries we start our RG calculation.  The
incompressible MHD equations in the Fourier space are
\eqnMHDubk
with
\begin{eqnarray}
P^{+}_{ijm}({\bf k}) & = & k_j P_{im}({\bf k}) + k_m P_{ij}({\bf k});
			\label{eqn:Pp} \\ 
P_{im}({\bf k}) & = & \delta_{im}-\frac{k_i k_m}{k^2}; \\
P^-_{ijm}({\bf k}) & = & k_j \delta_{im} - k_m \delta_{ij}; \label{eqn:Pm}\\
\hat{k} & = & ({\bf k},\omega); \\ d \hat{p} & = & d {\bf p} d
\omega/(2 \pi)^{d+1}.
\end{eqnarray}
Here $\nu$ and $\eta$ are the viscosity and the
resistivity respectively, and $d$ is the space dimension.

In our RG procedure the wavenumber range $(k_N,k_0)$ is divided
logarithmically into $N$ shells. The $n$th shell is $(k_n,k_{n-1})$ where $%
k_n=h^n k_0(h <1)$. In the following discussion,  we carry out the
elimination of the first shell $(k_1,k_0)$ and obtain the modified MHD\
equations. We then proceed iteratively to eliminate higher shells and get a
general expression for the modified MHD equations.
The renormalization group procedure is as follows:

\begin{enumerate}
\item  We divide the spectral space into two parts: 1.
the shell $(k_1,k_0)=k^>$, which is to be eliminated; 2.
$(k_N,k_1)=k^{<}$, set of modes to be retained.
Note that $\nu_0$ and $\eta_0$ denote
the viscosity and resistivity before the elimination of the first shell.

\item  We rewrite  Eqs.~(\ref{eqn:udot}, \ref{eqn:bdot}) for $k^{<}$ and 
$k^{>}$. The equations
for $u_i^{<}(\hat{k})$ and  $b_i^{<}(\hat{k})$ modes are
\bea
\left( -i\omega + \Sigma_{(0)}^{uu}(k) \right) u_i^<(\hat{k})  
+ \Sigma_{(0)}^{ub}(k)  b_i^<(\hat{k})  & =  &
-\frac{i}{2} P^+_{ijm}({\bf k}) \int d\hat{p}  (
      [u_j^< (\hat{p}) u_m^< (\hat{k}-\hat{p}) ]  \nonumber \\
& & + 2 [u_j^< (\hat{p}) u_m^> (\hat{k}-\hat{p}) ]
    + [u_j^> (\hat{p}) u_m^> (\hat{k}-\hat{p})] \nonumber \\
& & - \mbox{Similar terms for $b$} )
		\label{eqn:ukless} \\
\left( -i\omega + \Sigma_{(0)}^{bb}(k) \right) b_i^<(\hat{k})  
+ \Sigma_{(0)}^{bu}(k)  u_i^<(\hat{k})  & =  &
-i P^-_{ijm}({\bf k}) \int d\hat{p} (
      [u_j^< (\hat{p}) b_m^< (\hat{k}-\hat{p}) ] \nonumber \\
& & + [u_j^< (\hat{p}) b_m^> (\hat{k}-\hat{p}) 
    +u_j^> (\hat{p}) b_m^< (\hat{k}-\hat{p}) ] \nonumber \\
& & + [u_j^> (\hat{p}) b_m^> (\hat{k}-\hat{p})] )
		\label{eqn:bkless}
\eea
The $\Sigma$s appearing in the equations are usually called the
``self-energy'' in Quantum field theory language.  In the first
iteration, $\Sigma_{(0)}^{uu} = \nu_{(0)} k^2$ and $\Sigma_{(0)}^{bb}
= \eta_{(0)} k^2$, while the other two $\Sigma$s are zero.  The
equation for $u_i^> (\hat{k})$ modes can be obtained by interchanging
$<$ and $>$ in the above equations.

\item The terms given in the second and third brackets in the
Right-hand side of Eqs.~(\ref{eqn:ukless}, \ref{eqn:bkless}) are
calculated perturbatively.  Since we are interested in the statistical
properties of ${\bf u}$ and ${\bf b}$ fluctuations, we perform the
usual ensemble average of the system \cite{YakhOrsz}. We assume that
${\bf u}^{>}(\hat{k})$ and ${\bf b}^{>}(\hat{k})$ have gaussian
distributions with zero mean, while ${\bf u}^{<}(\hat{k})$ and ${\bf
b}^{<}(\hat{k})$ are unaffected by the averaging process.  Hence,

\bea
\left\langle u_i^> (\hat{k}) \right\rangle & = & 0 \label{eqn:avgbegin} \\ 
\left\langle b_i^> (\hat{k}) \right\rangle  & = & 0 \\ 
\left\langle u_i^< (\hat{k}) \right\rangle  & = & u_i^< (\hat{k}) \\
\left\langle b_i^< (\hat{k}) \right\rangle  & = & b_i^< (\hat{k})
\eea
and 
\begin{eqnarray}
\left\langle u_i^> (\hat{p}) u_j^> (\hat{q})\right\rangle  & = &
P_{ij}({\bf p)} C^{uu} (\hat{p}) \delta(\hat{p}+\hat{q}) \\
\left\langle b_i^> (\hat{p}) b_j^> (\hat{q})\right\rangle  &= &
P_{ij}({\bf p)} C^{bb} (\hat{p}) \delta(\hat{p}+\hat{q}) \\
\left\langle u_i^> (\hat{p}) b_j^> (\hat{q})\right\rangle  &= &
P_{ij}({\bf p)} C^{ub} (\hat{p}) \delta(\hat{p}+\hat{q}) \label{eqn:avgend}
\end{eqnarray}

The triple order correlations $\left\langle X_i^{>} (\hat{k}) X_j^{>}
(\hat{p}) X_m^{>} (\hat{q}) \right\rangle$ are zero due to Gaussian
nature of the fluctuations.  Here, $X$ stands for $u$ or $b$.  In
addition, we also neglect the contribution from the triple
nonlinearity $\la X_i^{<} (\hat{k}) X_j^{<} (\hat{p}) X_m^{<}
(\hat{q}) \ra$.  The last assumption is used in many of the turbulence
RG calculations \cite{YakhOrsz,McCo:book}.  Refer to Zhou and Vahala
\cite{ZhouVaha} for a discussion on the importance of the triple
nonlinearity.

\item The details of the perturbative calculation is given in Appendix
A.  To the first order, the second bracketed terms of
Eqs.~(\ref{eqn:ukless}, \ref{eqn:bkless}) vanish, but the nonvanishing
third bracketed terms yield corrections to $\Sigma$s.
Eqs.~(\ref{eqn:ukless}, \ref{eqn:bkless}) can now be approximated by
\bea 
\left( -i\omega + \Sigma_{(0)}^{uu} + 
	\delta \Sigma_{(0)}^{uu} \right) u_i^<(\hat{k}) + 
\left( \Sigma_{(0)}^{ub} + 
	\delta \Sigma_{(0)}^{ub} \right) b_i^<(\hat{k}) & = & 
-\frac{i}{2} P^+_{ijm}({\bf k}) \int d\hat{p}
	[u_j^< (\hat{p}) u_m^< (\hat{k}-\hat{p}) \nonumber \\ 
& & - b_j^<(\hat{p}) b_m^< (\hat{k}-\hat{p}) ] \\
\left( -i\omega + \Sigma_{(0)}^{bb} 
	+ \delta \Sigma_{(0)}^{bb} \right) b_i^<(\hat{k}) + 
\left( \Sigma_{(0)}^{bu} + 
	\delta \Sigma_{(0)}^{bu} \right) u_i^<(\hat{k}) & = & 
-i P^-_{ijm}({\bf k}) \int d\hat{p} 
[u_j^< (\hat{p}) b_m^< (\hat{k}-\hat{p}) ] \eea 
with 
\bea 
\delta \Sigma^{uu}_{(0)}(k) & = & \frac{1}{(d-1) k^2}
	\int_{\hat{p}+\hat{q}=\hat{k}}^{\Delta} d\hat{p} 
	[S(k,p,q) G^{uu}(\hat{p}) C^{uu}(\hat{q}) 
	-S_6(k,p,q) G^{bb}(\hat{p}) C^{bb}(\hat{q}) \nonumber \\ 
& & \hspace{1in} +S_6(k,p,q) G^{ub}(\hat{p}) C^{ub}(\hat{q})
	-S(k,p,q) G^{bu}(\hat{p}) C^{ub}(\hat{q})]  \label{eqn:sigmauu} \\
\delta \Sigma^{ub}_{(0)}(k) & = & \frac{1}{(d-1) k^2}
	\int_{\hat{p}+\hat{q}=\hat{k}}^{\Delta} d\hat{p} 
	[-S(k,p,q) G^{uu}(\hat{p})C^{ub}(\hat{q}) 
	+S_5(k,p,q) G^{ub}(\hat{p}) C^{uu}(\hat{q}) \nonumber \\ 
& & \hspace{1in} +S(k,p,q) G^{bu}(\hat{p}) C^{bb}(\hat{q})
	-S_5(k,p,q) G^{bb}(\hat{p}) C^{ub}(\hat{q})] \label{eqn:sigmaub} \\
\delta \Sigma^{bu}_{(0)}(k) & = & \frac{1}{(d-1) k^2}
	\int_{\hat{p}+\hat{q}=\hat{k}}^{\Delta} d\hat{p} 
	[S_8(k,p,q) G^{uu}(\hat{p}) C^{ub}(\hat{q}) 
	+S_{10}(k,p,q) G^{bb}(\hat{p}) C^{ub}(\hat{q}) \nonumber \\ 
& & \hspace{1in} +S_{12}(k,p,q) G^{ub}(\hat{p}) C^{bb}(\hat{q}) 
	-S_7(k,p,q) G^{bu}(\hat{p}) C^{uu}(\hat{q})] \label{eqn:sigmabu} \\ 
\delta \Sigma^{bb}_{(0)}(k) & = & \frac{1}{(d-1) k^2} 
	\int_{\hat{p}+\hat{q}=\hat{k}}^{\Delta} d\hat{p}
	[-S_8(k,p,q) G^{uu}(\hat{p}) C^{bb}(\hat{q}) 
	+S_9(k,p,q) G^{bb}(\hat{p}) C^{uu}(\hat{q}) \nonumber \\ 
& & \hspace{1in} +S_{11}(k,p,q) G^{ub}(\hat{p}) C^{ub}(\hat{q}) 
	-S_9(k,p,q)G^{bu}(\hat{p}) C^{ub}(\hat{q})] \label{eqn:sigmabb} 
\eea 
The quantities $S_i(k,p,q)$ are given in the Appendix~\ref{sec:Si}.
The integral $\Delta$ is to be done over the first shell.

The full-fledge calculation of $\Sigma$s is quite involved.
Therefore, we take two special cases: (1) $C^{ub}=0$ or $\sigma_c=0$;
and (2) $C^{ub} \approx C^{uu} \approx C^{ub}$ or $\sigma_c
\rightarrow 1$.  In this section we will discuss only the case
$\sigma_c=0$. The other case will be taken up in the next section.  A
word of caution is in order here.  In our calculation the parameters
used are $\sigma_c(k)=2 C^{ub}(k)/(C^{uu}(k)+C^{bb}(k))$ and $r_A(k) =
E^u(k)/E^b(k)$.  These parameters differ from the global $\sigma_c$
and $r_A$, yet we have restricted ourselves to the limiting cases of
constant $\sigma_c(k)$ and $r_A(k)$ because of simplicity.

When $\sigma_c=0$, an inspection of the self-energy diagrams shows
that $\Sigma^{ub}=\Sigma^{bu}=0$, and $G^{ub}=G^{bu}=0$.  Clearly, the
equations become much simpler because of the diagonal nature of matrices
$G$ and $\Sigma$.  The two quantities of interest $\delta
\Sigma_{(0)}^{uu}$ and $\delta \Sigma_{(0)}^{bb}$ will be given by
\bea \delta \Sigma^{uu}(\hat{k}) & = & \frac{1}{d-1}
				    \int_{\hat{p}+\hat{q}=\hat{k}}^{\Delta} 
				    d\hat{p} \left( 
					S(k,p,q) G^{uu}(p)C^{uu}(q) 
					- S_6(k,p,q) G^{bb}(p) C^{bb}(q) 
					\right) \\
\delta \Sigma^{bb}(\hat{k}) & = & \frac{1}{d-1}
				\int_{\hat{p}+\hat{q}=\hat{k}}^{\Delta}
				d\hat{p} \left( 
					-S_8(k,p,q) G^{uu}(p)C^{bb}(q) 
					+ S_9(k,p,q) G^{bb}(p)C^{uu}(q) 
					\right) 
\eea

The expressions for $\delta \Sigma$'s involves correlation functions
and Green's function, which are themselves functions of $\Sigma$'s.
In earlier RG calculations (see e.g. \cite{FNS,YakhOrsz}) the
correlation function $C(\hat{k})$ is function of forcing noise
spectrum.  In our self-consistent scheme we assume that we are in the
inertial range, and the energy spectrum is proportional to
Kolmogorov's $5/3$ powerlaw.  After that the renormalized Green's
function, or renormalized $\nu$ is computed iteratively.

The frequency dependence of the correlation function are taken as:
$C^{uu}(k,\omega)=2 C^{uu}(k) \Re(G^{uu}(k,\omega))$ and
$C^{bb}(k,\omega)=2 C^{bb}(k) \Re(G^{bb}(k,\omega))$.  In other words,
the relaxation time-scale of correlation function is assumed to be the
same as that of corresponding Green's function.  Since we are
interested in large time-scale behaviour of turbulence, we take the
limit $\omega$ going to zero.  Under these assumptions, the above
equations become
\bea 
\delta \nu_{(0)}(k) & = & \frac{1}{(d-1)k^2} 
			  \int^{\Delta}_{\bf p+q=k}
				 \frac{d {\bf p}}{(2 \pi)^d}
 			  [ S(k,p,q) \frac{C^{uu}(q)}
			  {\nu_{(0)}(p) p^2+\nu_{(0)}(q) q^2} 
			       \nonumber \\   
 		    & & \hspace{1.5in} - S_6(k,p,q) \frac{C^{bb}(q)}
			{\eta_{(0)}(p) p^2+\eta_{(0)}(q) q^2} ] 
                         \label{eqn:nu} \\
\delta \eta_{(0)}(k) & = & \frac{1}{(d-1)k^2}
			      \int^{\Delta}_{\bf p+q=k}
				 \frac{d {\bf p}}{(2 \pi)^d}
			      [ - S_9(k,p,q) \frac{C^{bb}(q)}
			      {\nu_{(0)}(p) p^2+\eta_{(0)}(q) q^2}
				\nonumber \\   
                        &  & \hspace{1.5in} +S_9(k,p,q) \frac{C^{uu}(q)}
			     {\eta_{(0)}(p) p^2+\nu_{(0)}(q) q^2} ] 
                             \label{eqn:eta}
\eea
Note that $\nu(k)=\Sigma^{uu}(k)/k^2$ and $\eta(k)=\Sigma^{bb}(k)/k^2$.

There are some important points to remember in the above step.
The frequency integral in the above is done
using contour integral. It can be shown that the integrals are nonzero 
only when
both the components appearing the denominator are of the same sign.
For example, first term of Eq.~(\ref{eqn:eta}) is nonzero only when
both $\nu_{(0)}(p)$ and $\eta_{(0)}(q)$ are of the same sign.

Let us denote $\nu_{(1)}(k)$ and $\eta_{(1)}(k)$ as the
renormalized viscosity and resistivity respectively after the first
step of wavenumber elimination.  Hence,
\begin{eqnarray}
\nu_{(1)} (k) & = & \nu_{(0)} (k) +\delta \nu_{(0)} (k); \\
\eta_{(1)} (k) & = & \eta_{(0)} (k) +\delta \eta_{(0)} (k) 
\label{eqn:eta_1}
\end{eqnarray}

\item  We keep eliminating the shells one after the other by the above
procedure. After $n+1$ iterations we obtain 
\begin{eqnarray}
\label{nu_n}
\nu_{(n+1)} (k)=\nu_{(n)} (k) +\delta \nu_{(n)} (k) \\
\eta_{(n+1)} (k)=\eta_{(n)} (k) +\delta \eta_{(n)} (k) 
\end{eqnarray}

where the equations for $\delta \nu_{(n)}(k)$ and $\delta
\eta_{(n)}(k)$ are the same as the Eqs.~(\ref{eqn:nu}, \ref{eqn:eta})
except that $\nu_{(0)}(k)$ and $\eta_{(0)}(k)$ appearing in the
equations are to be replaced by $\nu_{(n)}(k)$ and $\eta_{(n)}(k)$
respectively.  Clearly $\nu_{(n+1)}(k)$ and $\eta_{(n+1)}$(k) are the
renormalized viscosity and resistivity after the elimination of the
$(n+1)$th shell.

\item We need to compute $\delta \nu_{(n)}$ and $\delta
\eta_{(n)}$ for various $n$. These computations, however, require
$\nu_{(n)}$ and $\eta_{(n)}$.  In our scheme we solve these
equations iteratively.  In Eqs.~(\ref{eqn:nu}, \ref{eqn:eta})
we substitute $C(k)$ by one dimensional energy spectrum $E(k)$
\begin{eqnarray}
C^{uu}(k) = \frac{2 (2 \pi)^d}{S_d (d-1)} k^{-(d-1)} E^{u}(k) \\
C^{bb}(k) = \frac{2 (2 \pi)^d}{S_d (d-1)} k^{-(d-1)} E^{b}(k)
\end{eqnarray}
where $S_d$ is the surface area of $d$ dimensional spheres.  We assume
that $E^u(k)$ and $E^b(k)$ follow Eqs.~(\ref{eqn:Euk}, \ref{eqn:Ebk})
respectively.
Regarding $\nu_{(n)}$ and $\eta_{(n)}$, we attempt the following
form of solution
\bea
\nu_{(n)} (k_n k') & = & (K^u)^{1/2} \Pi^{1/3} k_n^{-4/3} \nu_{(n)}^* (k') \\
\eta_{(n)} (k_n k') & = & (K^u)^{1/2} \Pi^{1/3} k_n^{-4/3} \eta_{(n)}^* (k') 
\eea
with $k=k_{n+1}k' (k' < 1)$. We expect $\nu_{(n)}^*(k')$ and 
$\eta_{(n)}^*(k')$ to be  universal functions for large $n$.  
The substitution of $C^{uu}(k), C^{bb}(k), \nu_{(n)}(k)$, and
$\eta_{(n)}(k)$ yields the following equations:
\bea
\label{rg1}
\delta \nu_{(n)}^{*} (k') &  = & \frac{1}{(d-1)} 
				\int_{\bf p'+q'=k'} d{\bf q'}  
                         	\frac{2}{(d-1)S_d} \frac{E^{u}(q')}{q'^{d-1}} 
				\times \nonumber \\
& & \left[ S(k',p',q') \frac{1}{\nu_{(n)}^*(h p') p'^2 + \nu_{(n)}^*(h q') q'^2} 
          -S_6(k',p',q') \frac{r_A^{-1}}{\eta_{(n)}^*(h p') p'^2 
				+ \eta_{(n)}^*(h q') q'^2} \right]  \\
\delta \eta_{(n)}^{*} (k') & = & \frac{1}{(d-1)} 
				    \int_{\bf p'+q'=k'} d{\bf q'}  
                                  \frac{2}{(d-1)S_d} \frac{E^{u}(q')}{q'^{d-1}} 
					\times \nonumber \\
& & \left[-S_8(k',p',q') \frac{1}{\nu_{(n)}^*(h p') p'^2 + \eta_{(n)}^*(h q') q'^2} 
          +S_9(k',p',q') \frac{r_A^{-1}}
          {\eta_{(n)}^*(h p') p'^2 + \nu_{(n)}^*(h q') q'^2} \right]  \\
\nu^{*}_{(n+1)}(k') &  = & h^{4/3} \nu^{*}_{(n)} (h k') 
                           + h^{-4/3} \delta \nu^{*}_{(n)} (k')  \\
\eta^{*}_{(n+1)}(k') & = & h^{4/3} \eta^{*}_{(n)} (h k') 
                              + h^{-4/3} \delta \eta^{*}_{(n)} (h k')  
\eea
where the integrals in the above equations are performed iteratively 
over a region 
$1 \leq p',q' \leq 1/h$  with the constraint that ${\bf p'+q'=k'}$.
Fournier and Frisch \cite{FourFris} showed the above
volume integral in $d$ dimension to be
\be
\int_{\bf p'+q'=k'} d{\bf q'}  = S_{d-1} \int dp dq 
	\left( \frac{p q}{k} \right)^{d-2} (\sin \alpha)^{d-3}
\ee
where $\alpha$ is the angle between vectors ${\bf p}$ and ${\bf q}$.

\item Now we solve the above four equations self consistently for
various $r_A$s.  We have taken $h=0.7$.  We start with constant values
of $\nu^{*}_{(0)}$ and $\eta^{*}_{(0)}$ and compute the integrals
using Gauss quadrature technique.  Once $\delta \nu^*_{(0)}$ and $\delta
\eta^*_{(0)}$ have been computed, we can calculate $\nu^{*}_{(1)}$ and
$\eta^*_{(1)}$.  We iterate this process till $\nu^{*}_{(m+1)}(k')
\approx \nu^{*}_{(m)}(k')$ and $\eta^{*}_{(m+1)}(k') \approx
\eta^{*}_{(m)}(k')$, that is, till they converge.  We have reported
the limiting $\nu^*$ and $\eta^*$ whenever the solution converges.
The criterion for convergence is that the error must be less than
1\%.  This criterion is usually achieved by $n=10$ or so.  The result
of our RG analysis is given below.

\end{enumerate}

\begin{figure}
\centerline{
        \psfig{figure=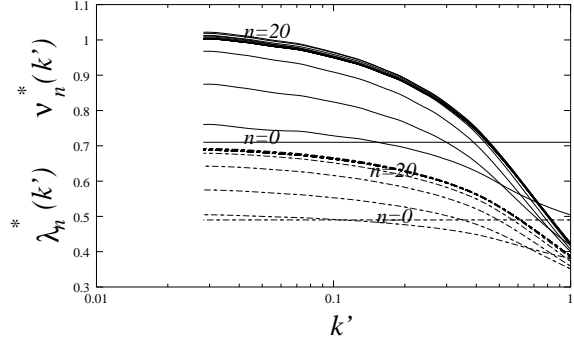,width=12cm,angle=0}}
        \vspace*{0.5cm}
\caption{Plot of $\nu^*(k')$ (solid) and $\eta^*(k')$ (dashed)
vs. $k'$ for $d=3$ and $\sigma_c=0, r_A=1$. Values at various
iterations are shown by different curves.}
\label{fig:d3ra1}
\end{figure}

\begin{figure}
\centerline{
        \psfig{figure=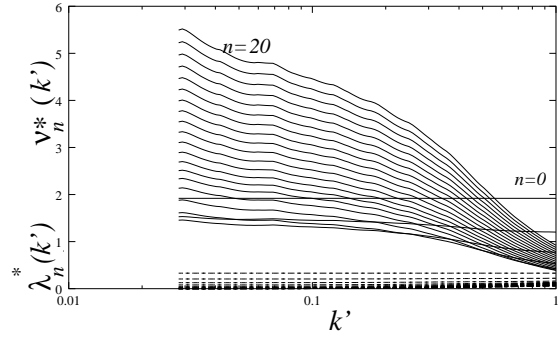,width=12cm,angle=0}}
        \vspace*{0.5cm}
\caption{Plot of $\nu^*(k')$ and $\eta^*(k')$ vs. $k'$ for
$d=2$ and $\sigma_c=0, r_A=1$.}
\label{fig:d2}
\end{figure}

\begin{figure}
\centerline{
        \psfig{figure=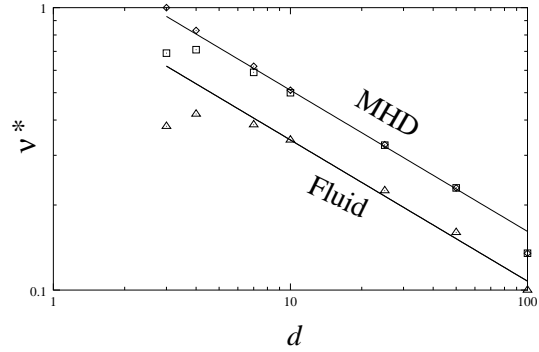,width=12cm,angle=0}}
        \vspace*{0.5cm}
\caption{Plot of asymptotic $\nu^*$ (square) and $\eta^*$ (diamond)
vs. $d$ for $\sigma_c=0$ and $r_A=1$.  The fluid $\nu^*$ (triangle) is
also plotted for reference.  The solid lines are the $d^{-1/2}$ curves
.}
\label{fig:nuvsd}
\end{figure}

We have carried out the RG analysis for various space dimensions and
find that the solution converges for all $d > d_c \approx 2.2$.
Hence, the RG fixed-point for MHD turbulence is stable for $d \ge
d_c$.  For illustration of convergent solution, see
Fig.~\ref{fig:d3ra1} for the plots of $\nu^{*}_{(n)}(k')$ and
$\eta^*_{(n)}(k')$ for $d=3, r_A=1$.  The RG fixed point for $d < d_c$
is unstable. Refer to the plots of Fig.~\ref{fig:d2} for $d=2, r_A=1$
as an example of an unstable solution. From this observation we can
claim that Kolmogorov's powerlaw is a consistent solution of MHD RG
equations at least for $d \ge d_c$.  The values of asymptotic ($k'
\rightarrow 0$ limit) $\nu^*$ and $\eta^*$ for various $d$ and $r_A=1$
are displayed in Table \ref{tab:alldra1}.  Clearly $\nu^*$ and
$\eta^*$ decreases with the increase in $d$ for $d \ge 4$.  The plot
of Fig.~(\ref{fig:nuvsd}) shows that both $\nu^*$ and $\eta^*$ are
proportional to $d^{-1/2}$.  In the same plot we have also displayed
$\nu^*$ for pure fluid turbulence; there too $\nu^* \propto d^{-1/2}$.
This result is consistent with Fournier {\em et al.}'s prediction for
fluid turbulence \cite{Four:inf}.

An important point to note is that the stability of RG fixed point in
a given space dimension depends on Alfv\'{e}n ratio and normalized
cross helicity.  For example, for $d=2.2$ the RG fixed point is stable
for $r_A \ge 1$, but unstable for $r_A < 1$.  A detailed study of
stability of the RG fixed point is required to ascertain the boundary
of stability.

The values of renormalized parameters for $d=3$ and various $r_A$ are
shown in Table \ref{tab:d3}.  For large $r_A$ (fluid dominated
regime), $\nu^*$ is close to renormalized viscosity of fluid
turbulence ($r_A = \infty$), but $\eta^*$ is also finite.  That
means there is a positive magnetic energy flux even when $r_A
\rightarrow \infty$.  As $r_A$ is decreased, $\eta^*$ decreases but
$\nu^*$ increases, or the Prandtl number $Pr=\nu/\eta$ increases. This
trend is seen till $r_A \approx 0.25$ where the RG fixed point with
nonzero $\nu^*$ and $\eta^*$ becomes unstable, and the trivial RG
fixed point with $\nu^*=\eta^*=0$ becomes stable.  This result
suggests an absence of turbulence for $r_A$ below 0.25 (approximately).
Note that in the $r_A \rightarrow 0$ (fully magnetic) limit, the MHD
equations become linear, hence there is no turbulence.  Surprisingly,
our RG calculation suggests that turbulence disappear near $r_A=0.25$
itself.

The final $\nu^*(k')$ and $\eta^*(k')$ are constant for small $k'$ but
shifts towards zero for larger $k'$ (see
Fig.~\ref{fig:d3ra1}). Similar behaviour has been seen by McComb and
coworkers \cite{McCoWatt} for fluid turbulence; this behaviour is
attributed to the neglect of triple nonlinearity, and the corrective
procedure has been prescribed by Zhou and Vahala \cite{ZhouVaha}.  For
simplicity we have not included the effects of triple nonlinearity in our
calculation.

Kraichnan's $k^{-3/2}$ energy spectrum [see Eq.~(\ref{eqn:Krai65})] and
$\sigma^{uu} = \sigma^{bb} \propto k B_0$ do not satisfy the RG equations.
This implies that Kraichnan's 3/2 powerlaw is not a consistent solution
of RG equations.

Pouquet \cite{Pouq:2D} and Ishizawa and Hattori \cite{Ishi:EDQNM}
calculated the contributions to the renormalized (eddy) viscosity and
resistivity from each of the four nonlinear terms of MHD equations
using EDQNM (Eddy-Damped-Quasi-Normal-Markovian) approximation.  This
calculation has been done for $d=2$.  Let us denote the quantities
$\nu^{uu}$, $\nu^{ub}$, $\eta^{bu}$, $\eta^{bb}$ as contributions from
the terms ${\bf u \cdot \nabla u}$, ${\bf -b \cdot \nabla b}$, ${\bf
-u \cdot \nabla b}$, ${\bf b \cdot \nabla u}$ respectively.  Note that
various cascade rates of MHD turbulence
\cite{Ishi:EDQNM,Ishi:flux,Dar:flux} can be inferred from these
parameters.  The cascade rate from inside of the $X$ sphere ($X<$) to
outside of the $Y$ sphere ($Y>$) is
\be 
\Pi^{X<}_{Y>} = \int_0^{k_N} 2 \nu^{XY}(k) k^2 E^{X}(k) dk
\ee
where $X,Y$ denote $u$ or $b$.  We have computed $\nu^{XY}$ for
$d \ge d_c$; their values are listed in Tables \ref{tab:alldra1} and
\ref{tab:d3}.  Our results show that $\eta^{bb}$ is positive.
Pouquet argues that $\eta^{bb}$ is negative, while Ishizawa finds
it to be positive.  Since our RG procedure is applicable only for $d
>d_c=2.2$, it is not proper to compare our results with those of
Pouquet \cite{Pouq:2D} and Ishizawa and Hattori \cite{Ishi:EDQNM}.
However, we can claim that the magnetic energy cascade rate
($\Pi^{b<}_{b>}$) is positive for all $d > d_c$ because
$\eta^{bb} > 0$.  

The eddy viscosity and resistivity listed in Table~\ref{tab:alldra1}
shows that $\eta^{bb*}$ and $\nu^{ub*}$ decrease as $d$ increases.  On
the contrary, $\eta^{bu*}$ and $\nu^{uu*}$ first increases till around
$d=5$ then decreases as $d$ increases.  Since MHD fluxes are
proportional to these parameters, the corresponding fluxes decrease as
$d$ is increased beyond 5. Near $d=2$, $\nu^{uu*}$ is negative,
reminiscent of inverse cascade energy in two-dimensional fluid
turbulence in the $5/3$ region.  Also, $\eta^{bu*}$ is negative near
$d=2$, i.e., there is a negative energy cascade from $b$-sphere to
outside $u$-sphere The above results are consistent with the recent
simulation results of Dar {\em et al.}~\cite{Dar:flux} on
two-dimensional MHD turbulence.

The variation of $\nu^{XY*}$ with $r_A$ for $d=3$ is displayed in
Tables~\ref{tab:d3}.  For large $r_A$, $\nu^{uu*}$ is close to the
fluid turbulence $\nu^*$. However, $\eta^{bb*}$ also finite. Hence
there is a significant magnetic energy cascade from $b$-sphere to
outside $b$-sphere even in fluid dominated case.  As $r_A$ is
decreased, $\nu^{ub*}$ and $\eta^{bu*}$ increase, hence cascade
rates from $u$-sphere to outside $b$-sphere, and $b$-sphere to outside
$u$-sphere start becoming important.  Incidently, both $\nu^{uu*}$ and
$\eta^{bb*}$ decrease as $r_A$ decreases.  

In this section we have calculated the renormalized viscosity and
resistivity for $\sigma_c=0$.  In the next section we will calculate
these parameters for the $\sigma_{c} \rightarrow 1$ limit.

\section{RG Calculation: $\sigma_{c} \rightarrow 1$ }

The {\bf u} and {\bf b} correlation is very high when
$\sigma_c \rightarrow 1$.  For  this case it is best to work 
with Els\"{a}sser variables ${\bf z^{\pm} = u \pm b}$. 
In terms of Els\"{a}sser variables, $\la |z^+|^2 \ra
\gg \la z^-|^2 \ra$. These types of fluctuations have been observed 
in the solar wind near the Sun.  However, by the time the solar wind
approaches Earth, the normalized cross helicity is close to
zero.

In this section we will briefly discuss the RG treatment for the above
case.  For the following discussion we will denote $\la
|z^-|^2\ra/\la|z^+|^2\ra = r= (1-\sigma_c)/(1+\sigma_c)$.  Clearly $r
\ll 1$.

MHD equations in terms of Els\"{a}sser variables are
\be
\label{eqn:MHDzk}
\left( -i\omega + \nu_{(0)\pm \pm} k^2 \right) z_i^{\pm}(\hat{k}) 
+ \nu_{(0) \pm \mp} k^2 z_i^{\mp}(\hat{k}) =
-i M_{ijm}({\bf k}) \int d\hat{p} 
[ z_j^{\mp }(\hat{p}) z_m^{\pm}(\hat{k}-\hat{p}) ] 
\ee
where 
\be
M_{ijm}({\bf k})= k_j P_{im}({\bf k})  \label{eqn:M}
\ee

Note that the above equations contain four dissipative coefficients
$\nu_{\pm \pm}$ and $\nu_{\pm \mp}$ instead of usual two constants
$\nu_{\pm}=(\nu \pm \eta)/2$.  This is because $+-$ symmetry is broken
in this case, consequently RG generates the other two constants.  We
carry out the same procedure as outlined in the previous section.
After $n+1$ steps of the RG calculation, the above equations become
\bea
\left[ -i\omega \mp \left( \nu_{(n) \pm \pm}(k)
       +\delta \nu_{(n) \pm \pm }(k)\right) k^2 \right] z_i^{\pm <}(\hat{k}) 
                & & \nonumber \\
       + \left( \nu_{(n) \pm \mp}(k)
            +\delta \nu_{(n) \pm \mp }(k)\right) k^2 z_i^{\mp <}(\hat{k}) & = & 
   -i M_{ijm}({\bf k}) \int d \hat{p} 
       z_j^{\mp <}(\hat{p}) z_m^{\pm <}(\hat{k}-\hat{p})
\eea
with
\bea
\delta \nu_{(n) ++}(k) & = & \frac{1}{(d-1) k^2} 
                             \int_{\hat{p}+\hat{q}=\hat{k}}^{\Delta} d\hat{p}
                             [S_1(k,p,q) G_{(n)}^{++}(\hat{p})  C^{--}(\hat{q})
                             +S_2(k,p,q) G_{(n)}^{+-}(\hat{p})  C^{--}(\hat{q})
				\nonumber  \\
                       &   & \hspace{1.5in}
			     +S_3(k,p,q) G_{(n)}^{-+}(\hat{p})  C^{+-}(\hat{q})
                             +S_4(k,p,q) G_{(n)}^{--}(\hat{p})  C^{+-}(\hat{q})] 
				\label{eqn:nupp} \\
\delta \nu_{(n) +-}(k) & = & \frac{1}{(d-1) k^2} 
                             \int_{\hat{p}+\hat{q}=\hat{k}}^{\Delta} d\hat{p}
                             [S_1(k,p,q) G_{(n)}^{+-}(\hat{p})  C^{-+}(\hat{q})
                             +S_2(k,p,q) G_{(n)}^{++}(\hat{p})  C^{-+}(\hat{q}) 
				\nonumber \\
                       &   & \hspace{1.5in}
			     +S_3(k,p,q) G_{(n)}^{--}(\hat{p})
                             C^{++}(\hat{q}) +S_4(k,p,q)
                             G_{(n)}^{-+}(\hat{p}) C^{++}(\hat{q})]
           \label{eqn:nupm} 
\eea 
where the integral is performed over the $(n+1)$th shell
$(k_{n+1},k_n)$.  The equations for the other two $\delta \nu$s can be
obtained by interchanging $+$ and $-$ signs.  Now we assume that the
Alfv\'{e}n ratio is one, i.e., $C^{+-}=E^u - E^b = 0$.  Under this 
condition, the above equations reduce to
\bea
\delta \nu_{(n) ++}(k) & = & \frac{1}{(d-1) k^2} 
                             \int_{\hat{p}+\hat{q}=\hat{k}}^{\Delta} d\hat{p}
                             [S_1(k,p,q) G_{(n)}^{++}(\hat{p})  
                             +S_2(k,p,q) G_{(n)}^{+-}(\hat{p})] C^{--}(\hat{q})
	\label{eqn:nuppsig0} \\
\delta \nu_{(n) +-}(k) & = & \frac{1}{(d-1) k^2} 
                             \int_{\hat{p}+\hat{q}=\hat{k}}^{\Delta} d\hat{p}
             		     [S_3(k,p,q) G_{(n)}^{--}(\hat{p})
                             +S_4(k,p,q) G_{(n)}^{-+}(\hat{p})] C^{++}(\hat{q})
				\\
\delta \nu_{(n) -+}(k) & = & \frac{1}{(d-1) k^2} 
                             \int_{\hat{p}+\hat{q}=\hat{k}}^{\Delta} d\hat{p}
             		     [S_3(k,p,q) G_{(n)}^{++}(\hat{p})
                             +S_4(k,p,q) G_{(n)}^{+-}(\hat{p})] C^{--}(\hat{q})
				\\
\delta \nu_{(n) --}(k) & = & \frac{1}{(d-1) k^2} 
                             \int_{\hat{p}+\hat{q}=\hat{k}}^{\Delta} d\hat{p}
                             [S_1(k,p,q) G_{(n)}^{--}(\hat{p})  
                             +S_2(k,p,q) G_{(n)}^{-+}(\hat{p})] C^{++}(\hat{q})
	\label{eqn:nummsig0}
\eea 
The inspection of Eqs.~(\ref{eqn:nuppsig0}--\ref{eqn:nummsig0}) reveal
that $\nu_{++}$ and $\nu_{-+}$ are of the order of $r$.  Hence, we
take the $\hat{\nu}$ matrix to be of the form
\begin{equation}
\hat{\nu}(k,\omega) = \left(
\begin{array}{cc}
r \zeta & \alpha \\ 
r \psi  & \beta
\end{array}
\right)
\end{equation}

It is convenient to transform the frequency integrals in 
Eqs.~[(\ref{eqn:nuppsig0}-\ref{eqn:nummsig0})] into temporal integrals, which
leads 
\bea
\delta \nu_{(n) ++}(k) & = &  \frac{1}{(d-1) k^2} 
            \int_{\bf p+q=k}^{\Delta} \frac{d{\bf p}}{(2 \pi)^d}
	    \int_{-\infty}^{t} dt'
            [S_1(k,p,q) G_{(n)}^{++}({\bf p},t-t')  
            +S_2(k,p,q) G_{(n)}^{+-}({\bf p},t-t')] \nonumber \\
                       &   & \hspace{2.5in} C^{--}({\bf q},t-t')
\eea
and similar forms for equations for other $\nu$s.  Green's function
$\hat{G}(k,t-t') = \exp{- [\hat{\nu} k^2 (t-t')]}$ can be easily
evaluated by diagonalizing the matrix $\hat{\nu}$.  The final form of
$\hat{G}(k,t-t')$ to leading order in $r$ is
\be
\hat{G}(k,t-t') = \left(
\begin{array}{cc}
1-\frac{r \alpha \psi}{\beta^2} (1-e^{-\beta(t-t')}) &
- \left\{ \frac{\alpha}{\beta}+\frac{r \alpha}{\beta} 
         \left( \frac{\zeta}{\beta}- \frac{2 \alpha \psi}{\beta^2} \right)
  \right\}  (1-e^{-\beta (t-t')}) \\
-\frac{r \psi}{\beta} (1-e^{-\beta (t-t')}) &
e^{-\beta (t-t')} +\frac{r \alpha \psi}{\beta^2} 
			(1-e^{-\beta (t-t')}) 
\end{array}
\right)
\ee
The correlation matrix $\hat{C}(k,t-t')$ is given by
\bea
\left(
\begin{array}{cc}
C^{++}(k,t-t') & C^{+-}(k,t-t') \\ 
C^{-+}(k,t-t') & C^{--}(k,t-t') 
\end{array}
\right) 
& = & 
\hat{G}(k,t-t')
\left(
\begin{array}{cc}
C^{++}(k) & C^{+-}(k) \\ 
C^{-+}(k) & C^{--}(k) 
\end{array}
\right) 
\eea
The substitution of correlation functions and Green's functions yield
the following expressions for the elements of $\delta
\hat{\nu}$
\bea 
\delta \zeta_{(n)}(k) & = & \frac{1}{(d-1)k^2} \int^{\Delta} 
                          \frac{d {\bf p}}{(2 \pi)^d} C^+(q) 
\{  S_1(k,p,q) \frac{1}{\beta_{(n)}(q) q^2}  \nonumber \\
& & \hspace{1in}   + S_2(k,p,q) \frac{\alpha_{(n)}(p)}{\beta_{(n)}(p)} 
                    \left( \frac{1}{\beta_{(n)}(p)p^2+\beta_{(n)}(q) q^2} 
                     -\frac{1}{\beta_{(n)}(q) q^2} \right) \nonumber \\ 
& & \hspace{1in} - S_3(k,p,q) \frac{\alpha_{(n)}(q)}{\beta_{(n)}(q)} 
   \left( \frac{1}{\beta_{(n)}(p) p^2+\beta_{(n)}(q) q^2}
         -\frac{1}{\beta_{(n)}(p) p^2} \right) \} \\
\delta \alpha_{(n)}(k) & = & \frac{1}{(d-1)k^2} \int^{\Delta} \frac{d
	{\bf p}}{(2 \pi)^d} S_3(k,p,q) \frac{C^+(q)}{\beta_{(n)}(p)
	p^2} \\
\delta \psi_{(n)}(k) & = & \frac{1}{(d-1)k^2} \int^{\Delta} 
        \frac{d{\bf p}}{(2 \pi)^d} C^+(q) \{ 
          S_3(k,p,q) \frac{1}{\beta_{(n)}(q) q^2} \nonumber \\
& & \hspace{1in} + S_2(k,p,q) \frac{\alpha_{(n)}(q)}{\beta_{(n)}(q)} 
  \left(\frac{1}{\beta_{(n)}(p) p^2+\beta_{(n)}(q) q^2}
	-\frac{1}{\beta_{(n)}(p) p^2} \right) \nonumber \\ 
& & \hspace{1in} + S_4(k,p,q)\frac{\alpha_{(n)}(p)}{\beta_{(n)}(p)} 
                    \frac{1}{\beta_{(n)}(q)q^2} \} \\
\delta \beta_{(n)}(k) & = & \frac{1}{(d-1)k^2} \int^{\Delta} \frac{d
	{\bf p}}{(2 \pi)^d} S_1(k,p,q) \frac{C^+(q)}{\beta_{(n)}(p)
	p^2} 
\eea 

Note that $\delta \alpha$, $\delta \beta$, $\delta \zeta$, $\delta
\psi$, and hence $\alpha, \beta, \zeta, \psi$, are all independent of
$r$.  To solve the above equations we substitute the following
one-dimensional energy spectra in the above equations:
\bea 
E^{+}(k) & = & K^{+}
	\frac{\left( \Pi^{+} \right)^{4/3}} {\left( \Pi^{-}
	\right)^{2/3}} k^{-5/3} \\ 
E^-(k) & = & r E^+(k), 
\eea 
For the elements of $\hat{\nu}$  we substitute
\be 
	Z_{(n)}(k) = Z^*_{(n)}
	\sqrt{K^+} \frac{\left( \Pi^{+} \right)^{2/3}} {\left( \Pi^{-}
	\right)^{1/3}} k^{-4/3} 
\ee 
where $Z$ stands for $\zeta,\alpha,\psi,\beta$.  The renormalized
$Z^*$s are calculated using the procedure outlined in the previous
section.  For large $n$ their values for $d=3$ are
\begin{equation}
\hat{Z^*} = \left(
\begin{array}{cc}
0.86 r & 0.14 \\ 
0.16 r & 0.84
\end{array}
\right),
\end{equation}
and for $d=2$ they are
\begin{equation}
\hat{Z^*} = \left(
\begin{array}{cc}
0.95 r & 0.54\\ 1.10 r & 0.54
\end{array}
\right)
\end{equation}
Note that the solution converges for both $d=2$ and $d=3$.

As discussed in the earlier section, the cascade rates
$\Pi^{\pm}$ can be calculated from the renormalized
parameters discussed above.  Using the energy equations we
can easily derive the equations for the cascade rates,
which are
\bea 
\Pi^+ & = & \int_0^{k_N} 2 r \zeta k^2 E^+(k) + 
	    \int_0^{k_N} 2 \alpha k^2 (E^u(k)-E^b(k)) \\
\Pi^- & = & \int_0^{k_N} 2 \beta k^2 E^-(k) + 
	    \int_0^{k_N} 2 r \psi k^2 (E^u(k)-E^b(k)) 
\eea
Under the assumption that $r_A=1$, the parts of $\Pi^{\pm}$ 
proportional to $(E^u(k)-E^b(k))$ vanish. Hence, the total cascade
rate will be
\bea
\Pi & = & \frac{1}{2} (\Pi^+ + \Pi^-) \\
    & = & r \int_0^{k_N}  (\zeta+\beta) k^2 E^+(k)
\eea
Since $\zeta$ and $\beta$ are independent of $r$, the total cascade
rate is proportional to $r$ (for $r$ small).  Clearly the cascade
rate $\Pi$  vanishes when $r=0$ or $\sigma_c=1$.  This result is
consistent with the fact that the nonlinear interactions are absent
when only pure Alfv\'{e}n waves ($z^+$ or $z^-$) are present.  The
detailed calculation of the cascade rates $\Pi^{\pm}$ and the
constants $K^{\pm}$ is left for future studies (paper II).

\section{Summary and Conclusions}
 
In this paper we have constructed a self-consistent RG scheme for MHD
turbulence and computed the renormalized viscosity and resistivity.
These quantities find applications in simulations specially
large-eddy-simulations (LES).  In our RG scheme we assume that we are
in the fully nonlinear range, and the renormalized energy spectrum is
substituted for the correlation function.  After the substitution of
the correlation function, the renormalized Greens function or
renormalized viscosity and resistivity are computed iteratively.  In
our procedure Kolmogorov's powerlaw is taken for the energy spectrum,
and it is shown to be a self-consistent solution of the RG
equation for $d \ge d_c \approx 2.2$.  This result is consistent with
the recent theoretical \cite{MKV:B0RG,Srid1,Srid2}, numerical
\cite{MKV:mhdsim,Bisk1:Kolm,Bisk2:Kolm}, and observational studies of
solar wind \cite{MattGold,MarsTu90} that favor Kolmogorov's spectrum
for MHD turbulence.  Note that we do not carry out vertex (the
coefficient of the nonlinear term) renormalization; there is an
implicit assumption that the vertex renormalization, if any, is
included when we substitute renormalized energy spectrum
(Kolmogorov's) in the RG equation.

For simplicity of the calculation we have taken two special cases: (1)
$\sigma_c=0$ and full range of $r_A$; (2) $\sigma_c \rightarrow 1$ and
$r_A=1$.  For these two cases, the renormalized viscosity and
resistivity have been calculated for various space dimensions.  For
$\sigma_c=0$, there exists a stable RG fixed point for $d \ge d_c \approx
2.2$.  The RG fixed point is unstable for $d < d_c$.  Our result is
consistent with that Liang and Diamond \cite{Lian} where they conclude
that the RG fixed point for $d=2$ is unstable.  Note however that the
stability depends on the value of $r_A$ and $\sigma_c$.  An exhaustive
study is required to ascertain the boundary of stability as a function
of $d,\sigma_c$, and $r_A$.  

Some interesting trends emerge as we vary space dimensionality.
For $r_A=1$, the parameters $\nu^{ub*}$ and $\eta^{bb*}$ decrease
with the increase of $d$, but   $\nu^{uu*}$ and $\eta^{bu*}$
first increase till $d\approx 5$ then decrease. 

For $d=3$, variation of $r_A$ shows some interesting features.  For
large $r_A$, $\nu^*$ is close to the $\nu^*$ for pure fluid
turbulence, but $\eta^*$ is also finite for this case.  As $r_A$ is
decreased, $\nu^*$ increases and $\eta^*$ decreases, or $Pr=\nu/\eta$
increases.  Another important result of our calculation is absence of
turbulence for all $r_A$ below 0.25 or so.  In the $r_A \rightarrow 0$
(fully magnetic) limit, the MHD equations become linear, hence there
is no turbulence.  It is however surprising that turbulence disappears
near $r_A=0.25$ itself.

Our RG calculation for fluid turbulence ($r_A=\infty$) show several
interesting features.  The RG fixed point is stable for $d=2$, and the
renormalized viscosity is approximately $-0.5$, a {\em negative}
number.  This is consistent with the inverse cascade of energy for the
$5/3$ region.  Note however that for $d=3$, the renormalized fluid
viscosity is positive (0.38), hence the energy cascade is forward,
consistent with Kolmogorov's hypothesis.

The values of the renormalized viscosity and resistivity calculated by
our calculation differs from that of Verma and Bhattacharjee
\cite{MKVJKB}, the later calculation being similar to Kraichnan's
Direct-Interaction-Approximation calculation for fluid turbulence.
Verma and Bhattacharjee \cite{MKVJKB} assumed $\nu^{++}=\nu^{-+}$ and
$\nu^{+-}=\nu^{--}$, which is correct only in a limited region of
parameter space.  In addition, they assumed a cutoff for the
self-energy calculation to cure infrared divergence.  In the present
calculation we have overcome both these defects, and expect that the
numbers reported here will match with the future simulation results.
It is unfortunate that the RG scheme fails for $d=2$, hence, several
numerical results of 2D-MHD turbulence
\cite{MKV:mhdsim,Bisk1:Kolm,Bisk2:Kolm,Dar:flux} could not be compared
with the analytic results presented here.

Kraichnan's $3/2$ energy spectrum and $\Sigma^{uu} = \Sigma^{bb}
\propto (k B_0)$ do not satisfy renormalization group equations
[Eqs.~(\ref{eqn:nu}- \ref{eqn:eta_1})].  Hence $E(k) \propto k^{-3/2}$
is not a consistent solution of RG equations.  This result is in
agreement with the recent calculation by Verma \cite{MKV:B0RG}, where
it was shown that $B_0 = constant$ does not satisfy RG equations, but
the renormalized mean magnetic field ($B_0 \propto k^{-1/3}$) and
Kolmogorov's energy spectrum ($k^{-5/3}$) do satisfy the RG equations.
From these arguments we can claim that Kraichnan's $3/2$ powerlaw is
ruled out for strong MHD turbulence.  However, $k^{-3/2}$ may still be
considered in the framework of weak turbulence theory.

In our calculation we take $r_A(k)=E^u(k)/E^b(k)$ to be a constant.
The global Alfv\'{e}n ratio $E^u/E^b$ may differ significantly from
$r_A(k)$, still we make the approximation $r_A(k)=r_A$ to simplify the
calculation.  Similarly we have assumed that $\sigma_c(k) = const =
\sigma_c$.  In addition we also assume isotropy which does not hold in
the large $r_A$ limit.  It is hoped that future calculations will be
able to relax these assumptions and attempt solutions for more
realistic situations.

In many of the earlier RG calculations \cite{FNS,YakhOrsz} it is
assumed that the dynamical system under consideration is forced by
random noise at all scales. In those calculations, the noise and the
vertex (coefficient of the nonlinear term) usually get renormalized
along with the renormalization of dissipative coefficients.  In
contrast, in our scheme only the dissipative constants are
renormalized, and the renormalization correction to other parameters
(e.g. noise) are implicitly lumped into the full correlation function
by assuming Kolmogorov's powerlaw.  Our procedure is relatively
simple, and one can easily calculate various interesting
quantities. We hope that future developments in dynamical RG will be
able to justify some of the assumptions made in our procedure.

To conclude, we have applied RG procedure to MHD turbulence and
calculated the renormalized viscosity and resistivity.  These
parameters find applications in simulations.  We have shown that that
Kolmogorov's spectrum is a consistent solution of MHD RG equation. The
results presented here are in general agreement with the recent EDQNM
and numerical results. Validation of the results presented here with
numerical simulations will give us important insights into physics of
MHD turbulence.

\acknowledgements 
The author thanks J. K. Bhattacharjee for valuable discussions and
insights from the inception to the end of the problem.  The author
also benefitted greatly from the numerous discussions he had with
G. Dar and V. Eswaran on simulation results.

\appendix
\section{ Perturbative Calculation of MHD Equations}

The MHD equations (\ref{eqn:udot}, \ref{eqn:bdot}) can be written as
\bea 
\left(
\begin{array}{c}
u_i(\hat{k}) \\
b_i(\hat{k})
\end{array}
\right)
& = & 
\left(
\begin{array}{cc}
G^{uu}(\hat{k}) & G^{ub}(\hat{k})  \\
G^{bu}(\hat{k}) & G^{bb}(\hat{k}) 
\end{array}
\right) 
\left(
\begin{array}{cc}
-\frac{i}{2} P^+_{ijm}({\bf k}) 
\int d\hat{p} [ u_j (\hat{p}) u_m (\hat{k}-\hat{p}) -
   b_j (\hat{p}) b_m (\hat{k}-\hat{p}) ]  \\
- i P^-_{ijm}({\bf k}) 
\int d\hat{p} [ u_j (\hat{p}) b_m (\hat{k}-\hat{p}) ]
\end{array}
\right)
\eea
where the Greens function $G$ 
can be obtained from $G^{-1}(\hat{k})$ 
\begin{equation}
G^{-1}(k,\omega) = 
\left( 
\begin{array}{cc}
-i\omega - \Sigma^{uu}   & \Sigma^{ub}   \\
\Sigma^{bu}  & -i\omega - \Sigma^{bb}
\end{array}
\right) .
\end{equation}

We solve the above equation perturbatively keeping the
terms to the first order (nonvanishing).  As usual, we represent the
integrals by Feynmann diagrams. To the leading order,
the quantities $u_i$ and $b_i$ are expanded as 
\unitlength=1mm
\begin{fmffile}{fmfubfeyn}
\begin{equation}
\parbox{20mm}
{\begin{fmfgraph*}(20,15) 
  \fmfleft{i} \fmfright{o} \fmf{plain,label=$u_i$}{i,o}
\end{fmfgraph*}} = 
\parbox{20mm}
{\begin{fmfgraph*}(20,15) 
  \fmfleft{i} \fmfright{o1,o2} 
  \fmfv{d.sh=circle,d.f=full,d.si=0.1w}{v}
  \fmf{photon,label=$G^{uu}$}{i,v} 
  \fmf{plain,label=$u_m$,l.d=5}{v,o1}
  \fmf{plain,label=$u_j$,l.d=5}{v,o2}
\end{fmfgraph*}} - 
\parbox{20mm}
{\begin{fmfgraph*}(20,15) 
  \fmfleft{i} \fmfright{o1,o2} 
  \fmfv{d.sh=circle,d.f=full,d.si=0.1w}{v}
  \fmf{photon,label=$G^{uu}$}{i,v} 
  \fmf{dashes,label=$b_m$,l.d=5}{v,o1} 
  \fmf{dashes,label=$b_j$,l.d=5}{v,o2}
\end{fmfgraph*}} + 
\parbox{20mm}
{\begin{fmfgraph*}(20,15) 
  \fmfleft{i} \fmfright{o1,o2} 
  \fmfv{d.sh=circle,d.f=empty,d.si=0.1w}{v}
  \fmf{photon,label=$G^{ub}$,width=1}{i,v} 
  \fmf{dashes,label=$b_m$,l.d=5}{v,o1} 
  \fmf{plain,label=$u_j$,l.d=5}{v,o2}
\end{fmfgraph*}} + \ldots
\end{equation}
\begin{equation}
\parbox{20mm}
{\begin{fmfgraph*}(20,15) 
  \fmfleft{i} \fmfright{o} \fmf{dashes,label=$b_i$}{i,o}
\end{fmfgraph*}} =  
\parbox{20mm}
{\begin{fmfgraph*}(20,15) 
  \fmfleft{i} \fmfright{o1,o2} 
  \fmfv{d.sh=circle,d.f=full,d.si=0.1w}{v}
  \fmf{gluon,label=$G^{bu}$,width=1}{i,v} 
  \fmf{plain,label=$u_j$,l.d=5}{v,o2} 
  \fmf{plain,label=$u_m$,l.d=5}{v,o1}
\end{fmfgraph*}} - 
\parbox{20mm}
{\begin{fmfgraph*}(20,15) 
  \fmfleft{i} \fmfright{o1,o2} 
  \fmfv{d.sh=circle,d.f=full,d.si=0.1w}{v}
  \fmf{gluon,label=$G^{ub}$,width=1}{i,v} 
  \fmf{dashes,label=$b_j$,l.d=5}{v,o2} 
  \fmf{dashes,label=$b_m$,l.d=5}{v,o1}
\end{fmfgraph*}} + 
\parbox{20mm}
{\begin{fmfgraph*}(20,15) 
  \fmfleft{i} \fmfright{o1,o2} 
  \fmfv{d.sh=circle,d.f=empty,d.si=0.1w}{v}
  \fmf{gluon,label=$G^{bb}$}{i,v} 
  \fmf{plain,label=$u_j$,l.d=5}{v,o2} 
  \fmf{dashes,label=$b_m$,l.d=5}{v,o1}
\end{fmfgraph*}} +  \ldots
\end{equation}
\end{fmffile}
The variables $u$ and $b$ are represented by solid and dashed line 
respectively.  The quantity $G^{uu}, G^{ub}, G^{bb}, G^{bu}$ are represented
by thick wiggly (photon), thin wiggly, thick curly (gluon), and thin 
curly lines respectively.  The filled circle represents $(-i/2) P^+_{ijm}$
vertex, while the empty circle represents $-i P^-_{ijm}$
vertex.

The substitution of $u_i^>$ and $b_i^>$ in the second bracketed terms
of Eqs.~(\ref{eqn:ukless}, \ref{eqn:bkless}), denoted by $I_2^u$ and
$I_2^b$ respectively, will yield
\unitlength=1mm
\begin{fmffile}{fmfI2}
\begin{eqnarray}
I_2^u & = & 
 \parbox{20mm}
{\begin{fmfgraph*}(20,15) 
  \fmfleft{i} \fmfright{o1,o2} 
  \fmfv{d.sh=circle,l=2,l.a=180,l.d=.1w,d.f=full,d.si=0.1w}{v}
  \fmf{phantom}{i,v}
  \fmf{plain,label=$>$,l.d=5}{v,o2}
  \fmf{plain,label=$<$,l.d=5}{v,o1}
\end{fmfgraph*}} - 
 \parbox{20mm}
{\begin{fmfgraph*}(20,15) 
  \fmfleft{i} \fmfright{o1,o2} 
  \fmfv{d.sh=circle,l=2,l.a=180,l.d=.1w,d.f=full,d.si=0.1w}{v}
  \fmf{phantom}{i,v}
  \fmf{dashes,label=$>$,l.d=5}{v,o2}
  \fmf{dashes,label=$<$,l.d=5}{v,o1}
\end{fmfgraph*}}  \\
I_2^b & = &
 \parbox{20mm}
{\begin{fmfgraph*}(20,15) 
  \fmfleft{i} \fmfright{o1,o2} 
  \fmfv{d.sh=circle,d.f=empty,d.si=0.1w}{v}
  \fmf{phantom}{i,v}
  \fmf{plain,label=$>$,l.d=5}{v,o2}
  \fmf{dashes,label=$<$,l.d=5}{v,o1}
\end{fmfgraph*}} +
 \parbox{20mm}
{\begin{fmfgraph*}(20,15) 
  \fmfleft{i} \fmfright{o1,o2} 
  \fmfv{d.sh=circle,d.f=empty,d.si=0.1w}{v}
  \fmf{phantom}{i,v}
  \fmf{plain,label=$<$,l.d=5}{v,o2}
  \fmf{dashes,label=$>$,l.d=5}{v,o1}
\end{fmfgraph*}} 
\end{eqnarray}
We illustrate the expansion of one of the above diagrams:
\begin{eqnarray}
 \parbox{20mm}
{\begin{fmfgraph*}(20,15) 
  \fmfleft{i} \fmfright{o1,o2} 
  \fmfv{d.sh=circle,d.f=full,d.si=0.1w}{v}
  \fmf{phantom}{i,v}
  \fmf{plain,label=$>$,l.d=5}{v,o2}
  \fmf{plain,label=$<$,l.d=5}{v,o1}
\end{fmfgraph*}}
& = &
\parbox{20mm}
{\begin{fmfgraph*}(20,15) 
  \fmfv{d.sh=circle,l=2,l.a=180,l.d=0.1w,d.f=full,d.si=0.1w}{v1}
  \fmfv{d.sh=circle,d.f=full,d.si=0.1w}{v2}
  \fmfleft{i}  
  \fmf{phantom}{i,v1}
  \fmf{photon,tension=0.1,left}{v1,v2}
  \fmf{plain,right,label=$<>$}{v1,v2} 
  \fmfright{o}
  \fmf{plain,label=$>$,l.d=5}{v2,o} 
\end{fmfgraph*}}  +
\parbox{20mm}
{\begin{fmfgraph*}(20,15) 
  \fmfv{d.sh=circle,d.f=full,d.si=0.1w}{v1}
  \fmfv{d.sh=circle,d.f=full,d.si=0.1w}{v2}
  \fmfleft{i}  
  \fmf{phantom}{i,v1}
  \fmf{photon,tension=0.1,left}{v1,v2}
  \fmf{plain,right,label=$<>$}{v1,v2} 
  \fmfright{o}
  \fmf{plain,label=$<$,l.d=5}{v2,o} 
\end{fmfgraph*}}  +
\parbox{20mm}
{\begin{fmfgraph*}(20,15) 
  \fmfv{d.sh=circle,d.f=full,d.si=0.1w}{v1}
  \fmfv{d.sh=circle,d.f=full,d.si=0.1w}{v2}
  \fmfleft{i}  
  \fmf{phantom}{i,v1}
  \fmf{photon,tension=0.1,left}{v1,v2}
  \fmf{plain,right,label=$<<$}{v1,v2} 
  \fmfright{o}
  \fmf{plain,label=$>$,l.d=5}{v2,o} 
\end{fmfgraph*}}  +
\parbox{20mm}
{\begin{fmfgraph*}(20,15) 
  \fmfv{d.sh=circle,l=2,l.a=180,l.d=.1w,d.f=full,d.si=0.1w}{v1}
  \fmfv{d.sh=circle,d.f=full,d.si=0.1w}{v2}
  \fmfleft{i}  
  \fmf{phantom}{i,v1}
  \fmf{photon,tension=0.1,left}{v1,v2}
  \fmf{plain,right,label=$<<$}{v1,v2} 
  \fmfright{o}
  \fmf{plain,label=$<$,l.d=5}{v2,o} 
\end{fmfgraph*}} -  \nonumber \\
& & 
\parbox{20mm}
{\begin{fmfgraph*}(20,15) 
  \fmfv{d.sh=circle,label=2,l.a=180,l.d=.1w,d.f=full,d.si=0.1w}{v1}
  \fmfv{d.sh=circle,d.f=full,d.si=0.1w}{v2}
  \fmfleft{i}  
  \fmf{phantom}{i,v1}
  \fmf{photon,tension=0.1,left}{v1,v2}
  \fmf{dots,right,label=$<>$}{v1,v2} 
  \fmfright{o}
  \fmf{dashes,label=$>$,l.d=5}{v2,o} 
\end{fmfgraph*}}  -
\parbox{20mm}
{\begin{fmfgraph*}(20,15) 
  \fmfv{d.sh=circle,d.f=full,d.si=0.1w}{v1}
  \fmfv{d.sh=circle,d.f=full,d.si=0.1w}{v2}
  \fmfleft{i}  
  \fmf{phantom}{i,v1}
  \fmf{photon,tension=0.1,left}{v1,v2}
  \fmf{dots,right,label=$<>$}{v1,v2} 
  \fmfright{o}
  \fmf{dashes,label=$<$,l.d=5}{v2,o} 
\end{fmfgraph*}}  -
\parbox{20mm}
{\begin{fmfgraph*}(20,15) 
  \fmfv{d.sh=circle,d.f=full,d.si=0.1w}{v1}
  \fmfv{d.sh=circle,d.f=full,d.si=0.1w}{v2}
  \fmfleft{i}  
  \fmf{phantom}{i,v1}
  \fmf{photon,tension=0.1,left}{v1,v2}
  \fmf{dots,right,label=$<<$}{v1,v2} 
  \fmfright{o}
  \fmf{dashes,label=$>$,l.d=5}{v2,o} 
\end{fmfgraph*}}  -
\parbox{20mm}
{\begin{fmfgraph*}(20,15) 
  \fmfv{d.sh=circle,label=2,l.a=180,l.d=.1w,d.f=full,d.si=0.1w}{v1}
  \fmfv{d.sh=circle,d.f=full,d.si=0.1w}{v2}
  \fmfleft{i}  
  \fmf{phantom}{i,v1}
  \fmf{photon,tension=0.1,left}{v1,v2}
  \fmf{dots,right,label=$<<$}{v1,v2} 
  \fmfright{o}
  \fmf{dashes,label=$<$,l.d=5}{v2,o} 
\end{fmfgraph*}} + \nonumber \\
& & 
\parbox{20mm}
{\begin{fmfgraph*}(20,15) 
  \fmfv{d.sh=circle,d.f=full,d.si=0.1w}{v1}
  \fmfv{d.sh=circle,d.f=empty,d.si=0.1w}{v2}
  \fmfleft{i}  
  \fmf{phantom}{i,v1}
  \fmf{photon,width=1,tension=0.1,left}{v1,v2}
  \fmf{plain,right,label=$<>$}{v1,v2} 
  \fmfright{o}
  \fmf{dashes,label=$>$,l.d=5}{v2,o} 
\end{fmfgraph*}} +
\parbox{20mm}
{\begin{fmfgraph*}(20,15) 
  \fmfv{d.sh=circle,d.f=full,d.si=0.1w}{v1}
  \fmfv{d.sh=circle,d.f=empty,d.si=0.1w}{v2}
  \fmfleft{i}  
  \fmf{phantom}{i,v1}
  \fmf{photon,width=1,tension=0.1,left}{v1,v2}
  \fmf{plain,right,label=$<>$}{v1,v2} 
  \fmfright{o}
  \fmf{dashes,label=$<$,l.d=5}{v2,o} 
\end{fmfgraph*}} +
\parbox{20mm}
{\begin{fmfgraph*}(20,15) 
  \fmfv{d.sh=circle,d.f=full,d.si=0.1w}{v1}
  \fmfv{d.sh=circle,d.f=empty,d.si=0.1w}{v2}
  \fmfleft{i}  
  \fmf{phantom}{i,v1}
  \fmf{photon,width=1,tension=0.1,left}{v1,v2}
  \fmf{plain,right,label=$<<$}{v1,v2} 
  \fmfright{o}
  \fmf{dashes,label=$>$,l.d=5}{v2,o} 
\end{fmfgraph*}} + 
\parbox{20mm}
{\begin{fmfgraph*}(20,15) 
  \fmfv{d.sh=circle,d.f=full,d.si=0.1w}{v1}
  \fmfv{d.sh=circle,d.f=empty,d.si=0.1w}{v2}
  \fmfleft{i}  
  \fmf{phantom}{i,v1}
  \fmf{photon,width=1,tension=0.1,left}{v1,v2}
  \fmf{plain,right,label=$<<$}{v1,v2} 
  \fmfright{o}
  \fmf{dashes,label=$<$,l.d=5}{v2,o} 
\end{fmfgraph*}} +  \nonumber \\
& & 
\parbox{20mm}
{\begin{fmfgraph*}(20,15) 
  \fmfv{d.sh=circle,d.f=full,d.si=0.1w}{v1}
  \fmfv{d.sh=circle,d.f=empty,d.si=0.1w}{v2}
  \fmfleft{i}  
  \fmf{phantom}{i,v1}
  \fmf{photon,width=1,tension=0.1,left}{v1,v2}
  \fmf{dots,right,label=$<>$}{v1,v2} 
  \fmfright{o}
  \fmf{plain,label=$>$,l.d=5}{v2,o} 
\end{fmfgraph*}} +
\parbox{20mm}
{\begin{fmfgraph*}(20,15) 
  \fmfv{d.sh=circle,d.f=full,d.si=0.1w}{v1}
  \fmfv{d.sh=circle,d.f=empty,d.si=0.1w}{v2}
  \fmfleft{i}  
  \fmf{phantom}{i,v1}
  \fmf{photon,width=1,tension=0.1,left}{v1,v2}
  \fmf{dots,right,label=$<>$}{v1,v2} 
  \fmfright{o}
  \fmf{plain,label=$<$,l.d=5}{v2,o} 
\end{fmfgraph*}} +
\parbox{20mm}
{\begin{fmfgraph*}(20,15) 
  \fmfv{d.sh=circle,d.f=full,d.si=0.1w}{v1}
  \fmfv{d.sh=circle,d.f=empty,d.si=0.1w}{v2}
  \fmfleft{i}  
  \fmf{phantom}{i,v1}
  \fmf{photon,width=1,tension=0.1,left}{v1,v2}
  \fmf{dots,right,label=$<<$}{v1,v2} 
  \fmfright{o}
  \fmf{plain,label=$>$,l.d=5}{v2,o} 
\end{fmfgraph*}} +
\parbox{20mm}
{\begin{fmfgraph*}(20,15) 
  \fmfv{d.sh=circle,d.f=full,d.si=0.1w}{v1}
  \fmfv{d.sh=circle,d.f=empty,d.si=0.1w}{v2}
  \fmfleft{i}  
  \fmf{phantom}{i,v1}
  \fmf{photon,width=1,tension=0.1,left}{v1,v2}
  \fmf{dots,right,label=$<<$}{v1,v2} 
  \fmfright{o}
  \fmf{plain,label=$<$,l.d=5}{v2,o} 
\end{fmfgraph*}}  \nonumber \\
& & + \mbox{higer oder diagrams}
\end{eqnarray} 
\end{fmffile}   

In the above diagrams solid lines denote $C^{uu}$, and the dotted
lines denote $C^{ub}$.  The correlation function $C^{bb}$ is denoted
by dashed line.  As mentioned earlier, the wiggly and curly lines
denote various Green's functions.  All the diagrams except 4,8,12, and
16th can be shown to be trivially zero using
Eqs.~(\ref{eqn:avgbegin}---\ref{eqn:avgend}).  We assume that 4,8,12,
and 16th diagrams are also zero, as usually done in turbulence RG
calculations \cite{YakhOrsz,McCo:book}.  Similarly we can show that
all other diagrams of $I_2^u$ and $I_2^b$ are zero, hence, $I_2^{u}=
I_2^{b} = 0$ to first order.

The third bracketed terms of (\ref{eqn:ukless}, \ref{eqn:bkless}),
denoted by $I_3^{u}$ and $I_3^{b}$ respectively, are diagrammatically
represented as
\unitlength=1mm
\begin{fmffile}{fmfI3}
\begin{eqnarray}
I_3^u & = &
\parbox{20mm}
{\begin{fmfgraph*}(20,15) 
  \fmfleft{i} \fmfright{o1,o2} 
  \fmfv{d.sh=circle,d.f=full,d.si=0.1w}{v}
  \fmf{phantom}{i,v}
  \fmf{plain,label=$>$,l.d=5}{v,o2}
  \fmf{plain,label=$>$,l.d=5}{v,o1}
\end{fmfgraph*}} -
\parbox{20mm}
{\begin{fmfgraph*}(20,15) 
  \fmfleft{i} \fmfright{o1,o2} 
  \fmfv{d.sh=circle,d.f=full,d.si=0.1w}{v}
  \fmf{phantom}{i,v}
  \fmf{dashes,label=$>$,l.d=5}{v,o2}
  \fmf{dashes,label=$>$,l.d=5}{v,o1}
\end{fmfgraph*}} \nonumber \\
& = & - \delta \Sigma^{uu}(k)   
\parbox{20mm}
{\begin{fmfgraph*}(20,15) 
  \fmfleft{i} \fmfright{o}
  \fmf{plain,label=$<$,l.s=left}{i,o} 
\end{fmfgraph*}}  - \delta \Sigma^{ub}(k)   
\parbox{20mm}
{\begin{fmfgraph*}(20,15) 
  \fmfleft{i} \fmfright{o}
  \fmf{dashes,label=$<$,l.s=left}{i,o} 
\end{fmfgraph*}}  \\
I_3^b & = &
 \parbox{20mm}
{\begin{fmfgraph*}(20,15) 
  \fmfleft{i} \fmfright{o1,o2} 
  \fmfv{d.sh=circle,d.f=empty,d.si=0.1w}{v}
  \fmf{phantom}{i,v}
  \fmf{plain,label=$>$,l.d=5}{v,o2}
  \fmf{dashes,label=$>$,l.d=5}{v,o1}
\end{fmfgraph*}} \nonumber \\
& = & -\delta  \Sigma^{bu}(k)
\parbox{20mm}
{\begin{fmfgraph*}(20,15) 
  \fmfleft{i}
  \fmfright{o}
  \fmf{plain,label=$<$,l.s=left}{i,o} 
\end{fmfgraph*}} - \delta \Sigma^{bb}(k)
\parbox{20mm}
{\begin{fmfgraph*}(20,15) 
  \fmfleft{i}
  \fmfright{o}
  \fmf{dashes,label=$<$,l.s=left}{i,o} 
\end{fmfgraph*}}
\end{eqnarray}
where
\begin{eqnarray}
-(d-1) \delta \Sigma^{uu} & = &
\parbox{20mm}
{\begin{fmfgraph*}(20,15) 
  \fmfv{l=4,l.a=180,l.d=.1w,d.sh=circle,d.f=full,d.si=0.1w}{v1}
  \fmfv{d.sh=circle,d.f=full,d.si=0.1w}{v2}
  \fmfleft{i}  
  \fmf{phantom}{i,v1}
  \fmf{photon,tension=0.3,right}{v1,v2}
  \fmf{plain,tension=0.3,left,label=$>>$}{v1,v2} 
  \fmfright{o}   \fmf{phantom}{v2,o}
\end{fmfgraph*}} -
\parbox{20mm}
{\begin{fmfgraph*}(20,15) 
  \fmfv{l=2,l.a=180,l.d=.1w,d.sh=circle,d.f=full,d.si=0.1w}{v1}
  \fmfv{d.sh=circle,d.f=empty,d.si=0.1w}{v2}
  \fmfleft{i}  
  \fmf{phantom}{i,v1}
  \fmf{gluon,tension=0.3,right}{v1,v2}
  \fmf{dashes,tension=0.3,left,label=$>>$}{v1,v2} 
  \fmfright{o}   \fmf{phantom}{v2,o}
\end{fmfgraph*}} +
\parbox{20mm}
{\begin{fmfgraph*}(20,15) 
  \fmfv{l=2,l.a=180,l.d=.1w,d.sh=circle,d.f=full,d.si=0.1w}{v1}
  \fmfv{d.sh=circle,d.f=empty,d.si=0.1w}{v2}
  \fmfleft{i}  
  \fmf{phantom}{i,v1}
  \fmf{photon,width=1,tension=0.3,right}{v1,v2}
  \fmf{dots,tension=0.3,left,label=$>>$}{v1,v2} 
  \fmfright{o}   \fmf{phantom}{v2,o} 
\end{fmfgraph*}} -
\parbox{20mm}
{\begin{fmfgraph*}(20,15) 
  \fmfv{l=4,l.a=180,l.d=.1w,d.sh=circle,d.f=full,d.si=0.1w}{v1}
  \fmfv{d.sh=circle,d.f=full,d.si=0.1w}{v2}
  \fmfleft{i}  
  \fmf{phantom}{i,v1}
  \fmf{gluon,width=1,tension=0.3,right}{v1,v2}
  \fmf{dots,tension=0.3,left,label=$>>$}{v1,v2} 
  \fmfright{o}   \fmf{phantom}{v2,o} 
\end{fmfgraph*}}  \\
-(d-1) \delta \Sigma^{ub} & = & -
\parbox{20mm}
{\begin{fmfgraph*}(20,15) 
  \fmfv{l=4,l.a=180,l.d=.1w,d.sh=circle,d.f=full,d.si=0.1w}{v1}
  \fmfv{d.sh=circle,d.f=full,d.si=0.1w}{v2}
  \fmfleft{i}  
  \fmf{phantom}{i,v1}
  \fmf{photon,tension=0.3,right}{v1,v2}
  \fmf{dots,tension=0.3,left,label=$>>$}{v1,v2} 
  \fmfright{o}   \fmf{phantom}{v2,o}
\end{fmfgraph*}} +
\parbox{20mm}
{\begin{fmfgraph*}(20,15) 
  \fmfv{l=2,l.a=180,l.d=.1w,d.sh=circle,d.f=full,d.si=0.1w}{v1}
  \fmfv{d.sh=circle,d.f=empty,d.si=0.1w}{v2}
  \fmfleft{i}  
  \fmf{phantom}{i,v1}
  \fmf{photon,width=1,tension=0.3,right}{v1,v2}
  \fmf{plain,tension=0.3,left,label=$>>$}{v1,v2} 
  \fmfright{o}   \fmf{phantom}{v2,o}
\end{fmfgraph*}} +
\parbox{20mm}
{\begin{fmfgraph*}(20,15) 
  \fmfv{l=4,l.a=180,l.d=.1w,d.sh=circle,d.f=full,d.si=0.1w}{v1}
  \fmfv{d.sh=circle,d.f=full,d.si=0.1w}{v2}
  \fmfleft{i}  
  \fmf{phantom}{i,v1}
  \fmf{gluon,width=1,tension=0.3,right}{v1,v2}
  \fmf{dots,tension=0.3,left,label=$>>$}{v1,v2} 
  \fmfright{o}   \fmf{phantom}{v2,o} 
\end{fmfgraph*}} -
\parbox{20mm}
{\begin{fmfgraph*}(20,15) 
  \fmfv{l=2,l.a=180,l.d=.1w,d.sh=circle,d.f=full,d.si=0.1w}{v1}
  \fmfv{d.sh=circle,d.f=empty,d.si=0.1w}{v2}
  \fmfleft{i}  
  \fmf{phantom}{i,v1}
  \fmf{gluon,tension=0.3,right}{v1,v2}
  \fmf{dots,tension=0.3,left,label=$>>$}{v1,v2} 
  \fmfright{o}   \fmf{phantom}{v2,o} 
\end{fmfgraph*}} \\
-(d-1) \delta \Sigma^{bu} & = & 
\parbox{20mm}
{\begin{fmfgraph*}(20,15) 
  \fmfv{l=2,l.a=180,l.d=.1w,d.sh=circle,d.f=empty,d.si=0.1w}{v1}
  \fmfv{d.sh=circle,d.f=full,d.si=0.1w}{v2}
  \fmfleft{i}  
  \fmf{phantom}{i,v1}
  \fmf{photon,tension=0.3,left}{v1,v2}
  \fmf{dots,tension=0.3,right,label=$>>$}{v1,v2} 
  \fmfright{o}   \fmf{phantom}{v2,o}
\end{fmfgraph*}} +
\parbox{20mm}
{\begin{fmfgraph*}(20,15) 
  \fmfv{d.sh=circle,d.f=empty,d.si=0.1w}{v1}
  \fmfv{d.sh=circle,d.f=empty,d.si=0.1w}{v2}
  \fmfleft{i}  
  \fmf{phantom}{i,v1}
  \fmf{gluon,width=1,tension=0.3,right}{v1,v2}
  \fmf{dots,tension=0.3,left,label=$>>$}{v1,v2} 
  \fmfright{o}   \fmf{phantom}{v2,o}
\end{fmfgraph*}} +
\parbox{20mm}
{\begin{fmfgraph*}(20,15) 
  \fmfv{d.sh=circle,d.f=empty,d.si=0.1w}{v1}
  \fmfv{d.sh=circle,d.f=empty,d.si=0.1w}{v2}
  \fmfleft{i}  
  \fmf{phantom}{i,v1}
  \fmf{gluon,width=1,tension=0.3,left}{v1,v2}
  \fmf{dashes,tension=0.3,right,label=$>>$}{v1,v2} 
  \fmfright{o}   \fmf{phantom}{v2,o} 
\end{fmfgraph*}} -
\parbox{20mm}
{\begin{fmfgraph*}(20,15) 
  \fmfv{l=2,l.a=180,l.d=.1w,d.sh=circle,d.f=empty,d.si=0.1w}{v1}
  \fmfv{d.sh=circle,d.f=full,d.si=0.1w}{v2}
  \fmfleft{i}  
  \fmf{phantom}{i,v1}
  \fmf{gluon,width=1,tension=0.3,right}{v1,v2}
  \fmf{plain,tension=0.3,left,label=$>>$}{v1,v2} 
  \fmfright{o}   \fmf{phantom}{v2,o} 
\end{fmfgraph*}} \\
-(d-1) \delta \Sigma^{bb} & = & 
\parbox{20mm}
{\begin{fmfgraph*}(20,15) 
  \fmfv{l=2,l.a=180,l.d=.1w,d.sh=circle,d.f=empty,d.si=0.1w}{v1}
  \fmfv{d.sh=circle,d.f=full,d.si=0.1w}{v2}
  \fmfleft{i}  
  \fmf{phantom}{i,v1}
  \fmf{photon,tension=0.3,left}{v1,v2}
  \fmf{dashes,tension=0.3,right,label=$>>$}{v1,v2} 
  \fmfright{o}   \fmf{phantom}{v2,o}
\end{fmfgraph*}} +
\parbox{20mm}
{\begin{fmfgraph*}(20,15) 
  \fmfv{d.sh=circle,d.f=empty,d.si=0.1w}{v1}
  \fmfv{d.sh=circle,d.f=empty,d.si=0.1w}{v2}
  \fmfleft{i}  
  \fmf{phantom}{i,v1}
  \fmf{gluon,tension=0.3,right}{v1,v2}
  \fmf{plain,tension=0.3,left,label=$>>$}{v1,v2} 
  \fmfright{o}   \fmf{phantom}{v2,o}
\end{fmfgraph*}} +
\parbox{20mm}
{\begin{fmfgraph*}(20,15) 
  \fmfv{d.sh=circle,d.f=empty,d.si=0.1w}{v1}
  \fmfv{d.sh=circle,d.f=empty,d.si=0.1w}{v2}
  \fmfleft{i}  
  \fmf{phantom}{i,v1}
  \fmf{photon,width=1,tension=0.3,left}{v1,v2}
  \fmf{dots,tension=0.3,right,label=$>>$}{v1,v2} 
  \fmfright{o}   \fmf{phantom}{v2,o} 
\end{fmfgraph*}} -
\parbox{20mm}
{\begin{fmfgraph*}(20,15) 
  \fmfv{d.sh=circle,d.f=empty,d.si=0.1w}{v1}
  \fmfv{d.sh=circle,d.f=empty,d.si=0.1w}{v2}
  \fmfleft{i}  
  \fmf{phantom}{i,v1}
  \fmf{gluon,width=1,tension=0.3,right}{v1,v2}
  \fmf{dots,tension=0.3,left,label=$>>$}{v1,v2} 
  \fmfright{o}   \fmf{phantom}{v2,o} 
\end{fmfgraph*}} 
\end{eqnarray}
\end{fmffile}
%
In Eqs.~(A10--A13) we have omitted all the vanishing diagrams (similar
to those appearing in $I_2$).  Looking at the all the terms, we
observe $I_3^{u}$ contributes to $\Sigma^{uu}$ and $\Sigma^{ub}$
renormalization, while $I_3^{b}$ contributes $\Sigma^{bb}$ and
$\Sigma^{bu}$ renormalization.  The algebraic expressions for the
above diagrams are given by
Eqs.~(\ref{eqn:sigmauu}---\ref{eqn:sigmabb}).
 
\section{Values of $S_{\small i}$}
\label{sec:Si}

$S_i(k,p,q)$ are formed by contracting tensors $P^+_{ijm}, P^-_{ijm},
P_{ij}$ and $M_{ijm}$ (see Eqs.~(\ref{eqn:Pp},\ref{eqn:Pm},\ref{eqn:M})).
On simplification $S_i$s become functions of $k,p,q$ and cosines $x,y,z$
defined as
\be
{\bf p \cdot q} = -pqx; \hspace{1cm} {\bf q \cdot k}= qky; 
\hspace{1cm}{\bf p \cdot k}=pkz.
\ee
The algebraic expressions for various $S_i(k,p,q)$ are given
below.
\bea
S_1(k,p,q) & = & M_{bjm}(k) M_{mab}(p) P_{ja}(q) \nonumber \\
           & = & kp(d-2+z^2)(z+xy) \\
S_2(k,p,q) & = & M_{ajm}(k) M_{mab}(p) P_{jb}(q) \nonumber \\
           & = & kp(-z+z^3+y^2 z+ x y z^2) \\
S_3(k,p,q) & = & M_{bjm}(k) M_{jab}(p) P_{ma}(q) \nonumber \\
           & = & kp(-z+z^3+x^2 z+ x y z^2) \\
S_4(k,p,q) & = & M_{ajm}(k) M_{jab}(p) P_{mb}(q) \nonumber \\
           & = & kp(-z+z^3+x y+x^2 z+y^2 x+x y z^2)  \\
S(k,p,q)   & = & P^{+}_{bjm}(k) P^{+}_{mab}(p) P_{ja}(q) \nonumber \\
           & = & kp \left( (d-3)z+2 z^3+(d-1)xy \right) \\ 
S_5(k,p,q) & = & P^{+}_{bjm}(k) P^{-}_{mab}(p) P_{ja}(q) \nonumber \\
           & = & kp \left( (d-1)z+(d-3)xy-2 y^2 z \right) \\
S_6(k,p,q) & = & P^{+}_{ajm}(k) P^{-}_{mba}(p) P_{jb}(q) \nonumber \\
           & = & -S_5(k,p,q) \\
S_7(k,p,q) & = & P^{-}_{ijm}(k) P^{+}_{mab}(p) P_{ja}(q) P_{ib}(k) \nonumber \\
           & = & S_5(p,k,q) \\
S_8(k,p,q) & = & P^{-}_{ijm}(k) P^{+}_{jab}(p) P_{ma}(q) P_{ib}(k) \nonumber \\
           & = & -S_5(p,k,q) \\
S_9(k,p,q)    & = & P^{-}_{ijm}(k) P^{-}_{mab}(p) P_{ja}(q) P_{ib}(k) 
                           \nonumber \\
              & = & kp(d-1)(z+xy) \\
S_{10}(k,p,q) & = & P^{-}_{ijm}(k) P^{-}_{mab}(p) P_{jb}(q) P_{ia}(k) 
                           \nonumber \\
              & = & -S_9(k,p,q) \\
S_{11}(k,p,q) & = & P^{-}_{ijm}(k) P^{-}_{jab}(p) P_{ma}(q) P_{ib}(k) 
                           \nonumber \\
              & = & -S_9(k,p,q) \\
S_{12}(k,p,q) & = & P^{-}_{ijm}(k) P^{-}_{jab}(p) P_{mb}(q) P_{ia}(k) 
                           \nonumber \\
              & = & S_9(k,p,q) 
\eea

There are many useful relationships between $S_i(k,p,q)$s.  Some of 
them are
\bea
S(k,p,q) & = & S_1(k,p,q)+S_2(k,p,q)+S_3(k,p,q)+S_4(k,p,q)  \\
S_5(k,p,q) & = & S_1(k,p,q)-S_2(k,p,q)+S_3(k,p,q)-S_4(k,p,q)  
\eea


\newpage

\centerline{\large Figure Captions}
\vspace{1cm}

\noindent
{\bf Fig. 1.} \, Plot of $\nu^*(k')$ (solid) and $\eta^*(k')$ (dashed)
vs. $k'$ for $d=3$ and $\sigma_c=0, r_A=1$. Values at various
iterations are shown by different curves. \\

\noindent
{\bf Fig. 2.}  \, Plot of $\nu^*(k')$ and $\eta^*(k')$ vs. $k'$ for
$d=2$ and $\sigma_c=0, r_A=1$. \\

\noindent
{\bf Fig. 3.} \,  Plot of asymptotic $\nu^*$ (square) and $\eta^*$ (diamond) 
vs. $d$ for $\sigma_c=0$ and $r_A=1$.  The fluid $\nu^*$ (triangle) is also
plotted for reference.  The solid lines are the $d^{-1/2}$
curves. 

\newpage

\begin{table}
\caption{The values of $\nu^*,\eta^*,\nu^{uu*},\nu^{ub*},
\eta^{bu*},\eta^{bb*}$ for various space dimensions $d$ with
$r_A=1$ and $\sigma_c=0$.}
\label{tab:alldra1}
\begin{center} 
\begin{tabular}{lccccccr} 
 d  & $\nu^*$ & $\eta^*$ & $Pr$ & $\nu^{uu*}$ & $\nu^{ub*}$ 
		 &$\eta^{bu*}$ & $\eta^{bb*}$ \\ \hline
2.1 & $--$ & $--$ & $--$  & $--$  & $--$ & $--$ & $--$ \\
2.2 & 1.9  & 0.32 & 6.0   & -0.041& 1.96 & -0.44 & 0.76 \\ 
2.5 & 1.2  & 0.57 & 2.1   & 0.089 & 1.15 & -0.15 & 0.72 \\
3.0 & 1.00 & 0.69 & 1.4   & 0.20  & 0.80 & 0.078 & 0.61 \\
4.0 & 0.83 & 0.70 & 1.2   & 0.27  & 0.56 & 0.21  & 0.49 \\
7.0 & 0.62 & 0.59 & 1.1   & 0.26  & 0.36 & 0.25  & 0.34 \\
10.0& 0.51 & 0.50 & 1.0	  & 0.23  & 0.28 & 0.22  & 0.28 \\
50.0& 0.23 & 0.23 & 1.0   & 0.11  & 0.12 & 0.11  & 0.12 \\
100 & 0.14& 0.14& 1.0   & 0.065 & 0.069&0.066  & 0.069
\end{tabular}
\end{center}
\end{table}
\begin{table}
\caption{The values of $\nu^*,\eta^*,\nu^{uu*},\nu^{ub*},
\eta^{bu*},\eta^{bb*}$ for various $r_A$  when $ d=3$
and $\sigma_c=0$.}
\label{tab:d3}
\begin{center} 
\begin{tabular}{lccccccr} 
$r_A$  & $\nu^*$ & $\eta^*$ & $Pr$ & $\nu^{uu*}$ & $\nu^{ub*}$ 
		 &$\eta^{bu*}$ & $\eta^{bb*}$ \\ \hline
$\infty$ & 0.38 & $--$& $--$ & 0.38 & $--$ & $--$ & $--$ \\
5000 & 0.36 & 0.85 & 0.42  & 0.36 & 1.4E-4&-0.023 & 0.87 \\
100  & 0.36 & 0.85 & 0.42  & 0.36 & 7.3E-3&-0.022 & 0.87 \\
5    & 0.47 & 0.82 & 0.57  & 0.32 & 0.15 &-4.7E-4 & 0.82 \\
2    & 0.65 & 0.78 & 0.83  & 0.27 & 0.38 & 0.031  & 0.75 \\
1    & 1.00 & 0.69 & 1.4   & 0.20 & 0.80 & 0.078  & 0.61 \\
0.5  & 2.1 & 0.50 & 4.2   & 0.11 & 2.00 & 0.15   & 0.35 \\
0.3  & 11.0 & 0.14 & 78    & 0.022& 11.0 & 0.082  & 0.053 \\
0.2  & $--$ & $--$ & $--$ & $--$ & $--$ & $--$  \\
\end{tabular}
\end{center}
\end{table}

\end{document}